\documentclass[preprint]{aastex}
\usepackage{psfig,graphics}

\begin{document}

\shortauthors{Feigelson et al.}
\shorttitle{NGC 6334 Star Clusters}
\slugcomment{To appear in the Astronomical Journal}

\title{Stellar Clusters in the NGC 6334 Star Forming Complex}

\author{Eric D.  Feigelson, Amanda L. Martin, Collin J. McNeill, Patrick S. Broos, Gordon P. Garmire}

\affil{Department of Astronomy \& Astrophysics, The
Pennsylvania State University, 525 Davey Lab, University Park, PA
16802; {\tt edf@psu.edu}}

\begin{abstract}

The full stellar population of NGC 6334, one of the most spectacular regions of massive star formation in the nearby Galaxy, have not been well-sampled in past studies.  We analyze here a mosaic of two {\it Chandra X-ray Observatory} images of the region using sensitive data analysis methods, giving a list of 1607 faint X-ray sources with arcsecond positions and approximate line-of-sight absorption.  About 95\% of these are expected to be cluster members, most lower mass pre-main sequence stars.  Extrapolating to low X-ray levels, the total stellar population is estimated to be $20-30,000$ pre-main sequence stars.  The X-ray sources show a complicated spatial pattern with $\sim$10 distinct star clusters.  The heavily-obscured clusters are mostly associated with previously known far-infrared sources and radio HII regions.  The lightly-obscured clusters are mostly newly identified in the X-ray images.  Dozens of likely OB stars are found, both in clusters and dispersed throughout the region, suggesting that star formation in the complex has proceeded over millions of years.  A number of extraordinarily heavily absorbed X-ray sources are associated with the active regions of star formation. 

\end{abstract}

\keywords{ISM: clouds - ISM: individual (NGC~6334) - stars: early type - stars: formation - stars: pre-main sequence - X-Rays: stars}

\section{INTRODUCTION \label{intro.sec}}

The NGC~6634 (= Sharpless 8 = RCW 127) complex of molecular clouds and HII regions is an unusually rich region of massive star formation in the nearby Galaxy \citep[see review by][]{Persi08}. At a distance of $1.7$~kpc \citep{Neckel78} and Galactic coordinates (351.2+0.7), it exhibits many phenomena associated with embedded OB stars: luminous far-infrared sources denoting embedded clusters \citep{McBreen83}; radio continuum H~II regions \citep{Rodriguez82}; and complexes of molecular masers \citep{Moran80}.   Early studies argued that the star formation started in the center of the region, where the stellar cluster appears dynamically relaxed,  and proceeded outward where the embedded protostars are seen.  Recent high-resolution millimeter-band maps of the dust distribution reveal hundreds of molecular clumps distributed along a narrow ridge along the Galactic Plane \citep{Munoz07, Matthews08}.  The total molecular mass is estimated to be $\sim 2 \times 10^4$~M$_\odot$ with typical column density $\sim 5 \times 10^{22}$~cm$^{-2}$ along the ridge.  The total bolometric luminosity is $\sim 2 \times 10^6$ L$_\odot$ \citep{Loughran86} indicating that rich embedded stellar clusters with massive OB stars are present.  Optical and infrared maps are dominated by bright and spatially complex nebular emission from HII regions and heated dust.    

While deeply embedded massive stars in the region have been well-studied, less is known about the low mass pre-main sequence population associated with the massive stars.  $JHK$ source lists are incomplete due to heavy obscuration and H~II region nebulosity.  A wide field survey of $11 \la K \la 13.5$ sources finds hundreds of embedded low mass stars around the high mass regions \citep{Straw89b} and portions of the cloud have been surveyed to fainter levels in $JHK$ \citep{Tapia96, Persi00}.  At both bright and faint magnitudes, near-infrared samples are severely contaminated by unrelated Galactic field stars.  

High-resolution X-ray images of star forming molecular clouds from the {\it Chandra X-ray Observatory} can give rich samples of pre-main sequence stars which are complementary to those found in infrared surveys \citep[see review by][]{Feigelson07}.    The X-ray band reveals all types of young stars.  OB stars emit X-rays from their stellar winds on several scales: small instabilities in the wind acceleration region close to the stellar surface; magnetically channeled wind shocks;  colliding wind shocks of close binaries; and shocked winds on parsec scales suffusing large H~II regions \citep[e.g.][]{Cohen03, Townsley03, Schulz03, Gagne05, Skinner05, Sana06}.   Surveys of the Orion Nebula Cluster and Taurus clouds show that all lower mass stars emit X-rays during magnetic reconnection events near the stellar surface similar to, but far more powerful and frequent than, flares on the Sun \citep{Getman05, Gudel07}.  Studies with the high-resolution {\it Chandra X-ray Observatory} are particularly effective in producing stellar samples with several benefits: low contamination by unrelated older stars; penetration deep into the obscuring cloud; sampling of rich populations of lower mass stars; (sub)arcsecond positions; and selection that is mostly unbiased with respect to the presence or absence of protoplanetary disks.  {\it Chandra} studies of rich stellar clusters in obscured star forming regions around $1.5-2$~kpc distance include the Trifid Nebula \citep{Rho04},  RCW~38 \citep{Wolk06}, Cyg~OB2 \citep{Albacete07}, M~17 \citep{Broos07}, NGC~6357 near NGC~6334 \citep{Wang07}, and the Rosette Nebula and Molecular Cloud \citep{Wang08, Wang09}. 

Two {\it Chandra} images of NGC~6334 are presented by \citet{Ezoe06}.  Their study emphasizes the hard X-ray emission from individual stars and possibly from diffuse plasma suffusing the HII regions.  They extract a sample of 792 X-ray sources, but do not provide a source list to facilitate further study.  Prior to the {\it Chandra} observation, low resolution X-ray images gave only a glimpse of structure \citep{Sekimoto00}.  Ultra-hard X-ray emission attributed to background active galaxy has also been seen \citep{Bykov06}. 

We report here a reanalysis of the two {\it Chandra} exposures of NGC~6334 examined by \citet{Ezoe06} using advanced data analysis techniques.  Section~\ref{data_anal.sec} describes the data analysis methodology, \S\ref{sec:xcluster} gives the astronomical results, and \S\ref{discuss.sec} presents conclusions.  We find 1607 faint X-ray sources of which only $\sim$5\% are likely problem sources or non-cluster contaminants.  We give accurate positions and estimated absorption measurements for each star.  From this dataset, we obtain an improved view of the clustering in the region: 10 X-ray clusters are identified, including several previously unreported lightly obscured star clusters.  A considerable number of massive OB stars, some newly found, and deeply embedded protostars are detected.  These may include extremely young Class~0 massive protostars.

\section{X-RAY DATA ANALYSIS \label{data_anal.sec}}

\subsection{Chandra Observations \label{obs.sec}}

The X-ray observations discussed here were previously presented by \citet{Ezoe06}. NGC 6334 was observed in two overlapping exposures of the Imaging Array of the Advanced CCD Imaging Spectrometer (ACIS-I), each 40~ks in duration and covering a $\sim 17\arcmin \times 17\arcmin$ field of view. The {\it Chandra X-ray Observatory} satellite and ACIS instrument are described by \citet{Weisskopf02} and \citet{Garmire03}. The {\it Chandra} mirrors give an excellent subarcsecond point spread function (PSF) on-axis, but the PSF degrades substantially with off-axis distance. The field centers are $(\alpha,\delta) = (17^h20^m01.0^s,-35\arcdeg 56\arcmin 07\arcsec)$ and $(17^h20^m54.0^s, -35\arcdeg 47\arcmin 04\arcsec)$ with a satellite roll angle of 269\arcdeg.  The exposure times are 40.2~ks and 40.1~ks, respectively.  The observation was made in the standard Timed Exposure, Very Faint mode, with 3.2 s integration time and 5 pixel x 5 pixel event islands.  Pre-{\it Chandra} X-ray observations with the low-resolution $ASCA$ satellite are presented by \citet{Sekimoto00}.  They and \citet{Ezoe06} examine the spatially-integrated X-ray spectra of different components of NGC~6334 which we do not address here.  

\subsection{X-ray Source Identification \label{srcID.sec}}

\begin{figure}
\centering
\includegraphics[width=5.4in]{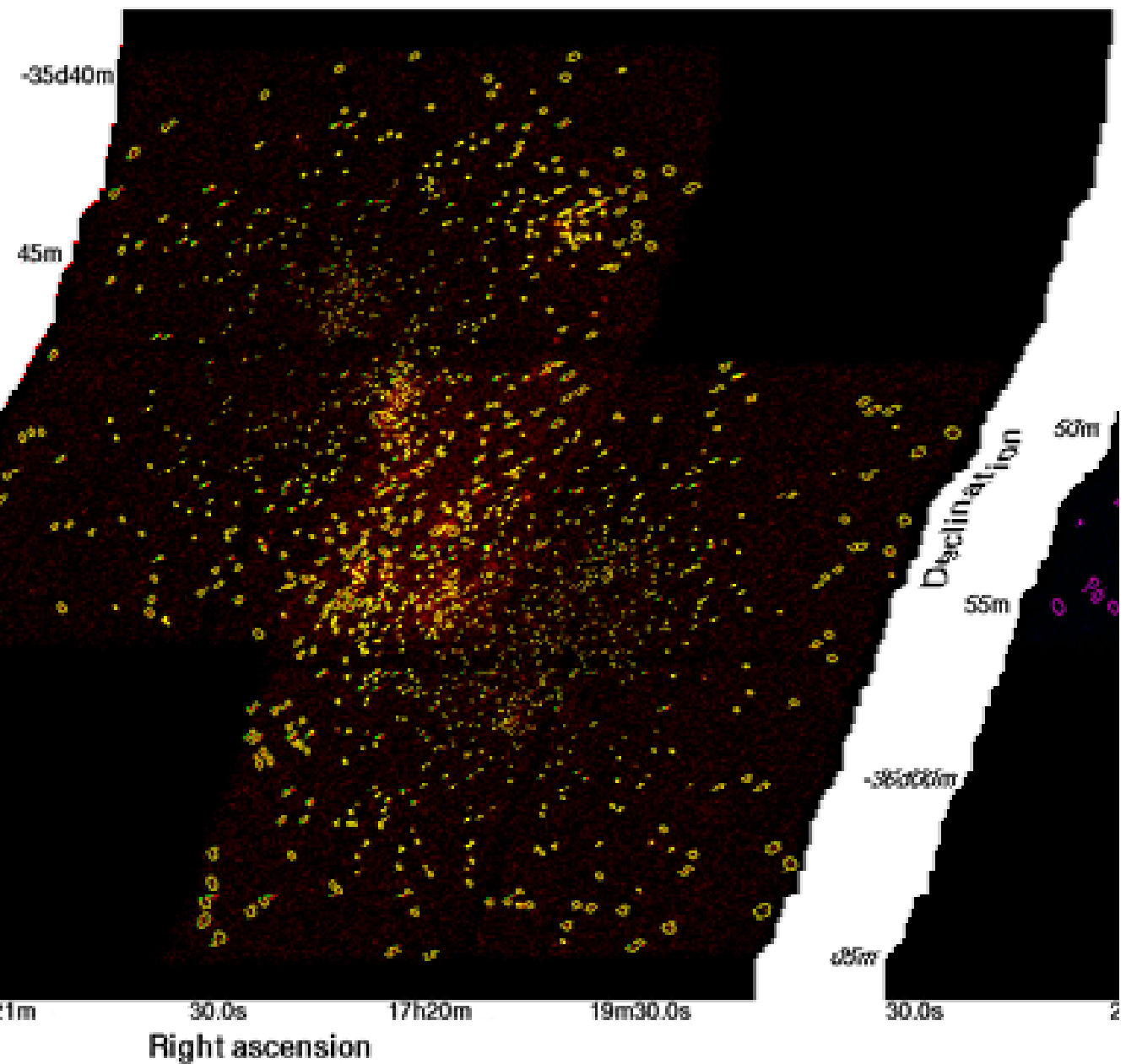}
\caption{Mosaic of two ACIS-I images of the {\it Chandra} of NGC~6334 with 1607 source extraction regions shown in yellow.   \label{fig:ACIS}}
\end{figure}

\begin{figure}[htbp]
\begin{center}
\includegraphics[width=5in]{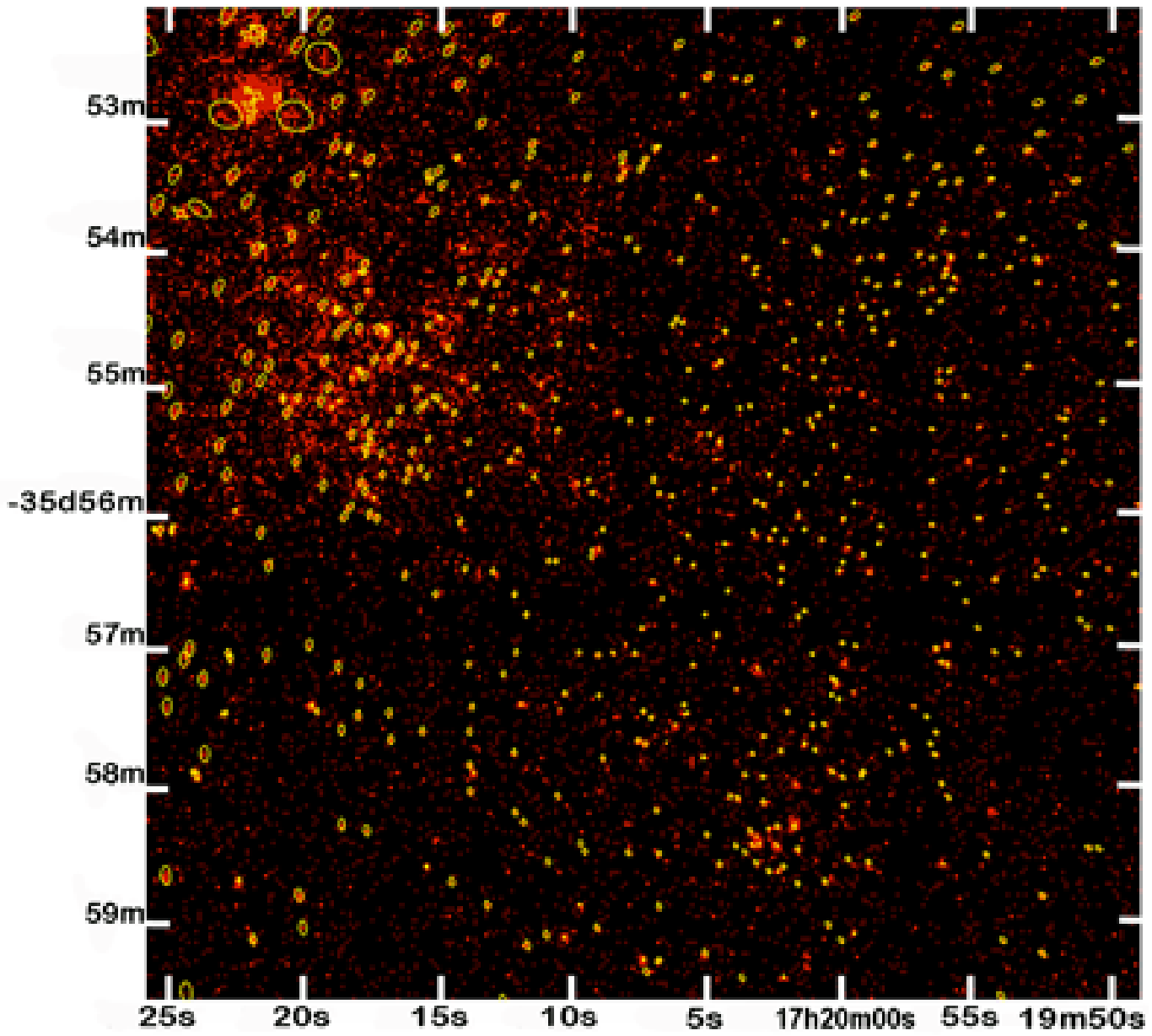}
\includegraphics[width=5in]{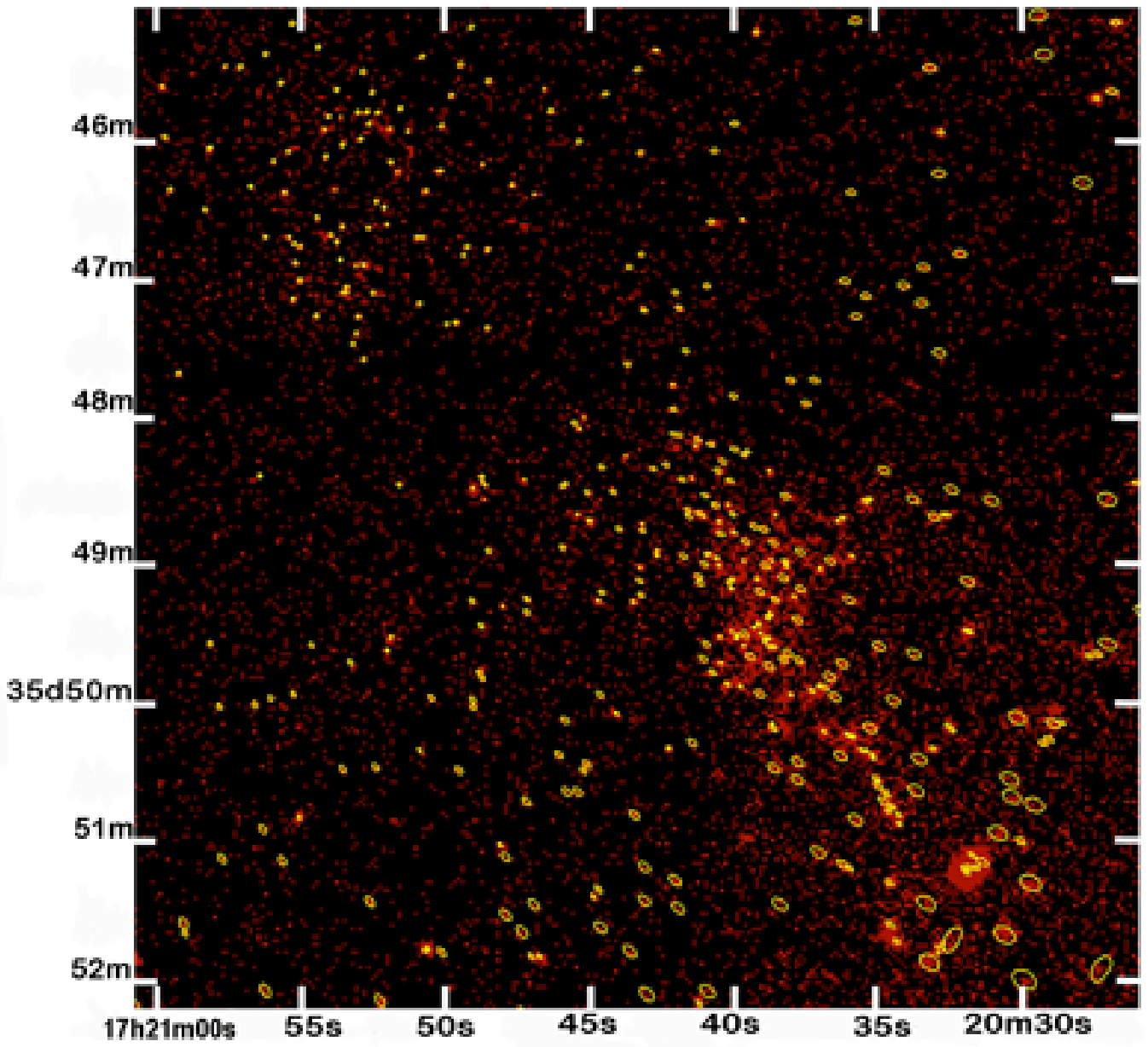}
\caption{Two expanded views of crowded portions of the {\it Chandra} fields. Top: Far-infrared regons NGC~6334 IV and V and the new clusters XA-XD.  Bottom: Far-infrared NGC~6334 I and II regions and the new cluster XJ.  See Figure~\ref{fig:space2} for the cluster identifications.}
\label{fig:expanded}
\end{center}
\end{figure}

Our data reduction procedure follows those of \citet{Townsley06a} and \citet{Broos07} with a few refinements. Level~1 event lists are first processed to recover an improved Level~2 event list as described by \citet[][Appendix~B]{Townsley03}. This includes filtering of bad events, removal of the $\pm 0.25$\arcsec\/ positional randomization, and the correction of charge transfer inefficiency. Cosmic-ray afterglows are removed by searching for multiple events arriving on the same detector pixel within a few CCD readout frames with decreasing energies. Sub-pixel positioning of events with non-zero CCD `grades' are refined using the algorithm of \citep{Mori01}. Matching bright X-ray sources in the central region of the detectors to stars in the 2MASS Point Source Catalog, a 0.35\arcsec\/ correction was applied to astrometrically align the fields to the {\it Hipparcos} reference frame. Figure~\ref{fig:ACIS} shows the resulting merged ACIS fields at reduced resolution, and Figure~\ref{fig:expanded} shows two expanded views. 

It is challenging to reliably identify faint sources in overlapping, crowded fields with a spatially varying PSF, and to establish their statistical significance. We adopt a procedure where a superset of possible sources is first constructed by a variety of means, following by iterative pruning of low-significance sources. An initial list was obtained by running the {\it wavdetect} wavelet detection algorithm \citep{Freeman02} with a low threshold of $10^{-5}$. Other potential sources are added from visual inspection of the images and from a peak-finding algorithm applied to a maximum-likelihood reconstruction of the images \citep{Lucy74} using local PSFs, as described by \citet{Townsley06a}. The Lucy-Richardson reconstruction is applied to the crowded central arcminute of the NGC~6334 fields with 0.1\arcsec\/ pixels. These steps are particularly important to identify closely-spaced sources which are merged by the wavelet algorithm. These procedures produced a superset of 2511 candidate sources.  A few sources missed by the algorithm found by visual examination were added.  

Photons are then extracted in polygonal regions closely matched to the local PSFs for each candidate source using the IDL-based program {\it ACIS Extract}\footnote{{\it ACIS Extract} ({\it AE}) and its associated Tools for ACIS Review and Analysis ({\it TARA}), developed by P. Broos and colleagues by the ACIS Team at Penn State, are publicly available at \url{http://www.astro.psu.edu/xray/docs/TARA}. AE is a versatile script written in IDL that performs source extraction, facilitates fitting X-ray spectra, creates light curves, and computes a wide variety of statistical properties of the sources. This analysis was performed using AE version 3.79.}. The number of extracted events in the total {\it Chandra} $0.5-8$~keV band is compared to the local source-free background level. The statistic $P_B$ measures the probability that the extracted events would be found from random Poisson fluctuations in the local background \citep[see the Appendix of][]{Weisskopf07}. This use of $P_B$ as a source existence criterion has the advantage of being independent of the poorly understood sensitivity limits of the {\it wavdetect} and maximum-likelihood restoration processes in the presence of crowding and spatially varying PSFs.

For our analysis here of the NGC~6634 fields, we choose a significance threshold of source existence $P_B < 0.001$ ($\log P_B < -3.0$) for source existence.  A list of candidate sources with $P_B<0.01$ was constructed to obtain measurements of local source-free background levels for the Poisson probability calculation.  For each source, $P_B$ values are calculated in three X-ray bands ($0.5-2$~keV, $2-8$~keV and $0.5-8$~keV) and, in the region overlapping the two ACIS fields, in both fields, and the lowest value is kept.   After four iterations of this procedure in {\it ACIS Extract}, where candidate source significances are recalculated based on revised background levels, we emerge with 1607 sources with $P_B < 0.001$. These sources are listed in Table~\ref{tbl:catalog_stub}. Note that we obtain twice as many sources as the 792 sources obtained by \citet{Ezoe06} using traditional data analysis methods.

Table~\ref{tbl:catalog_stub} has a format similar to the source tables in \citet{Getman05} and \citet{Townsley06a}. Column (1) gives the ACIS running sequence number, and column (2) provides the IAU designation. Columns (3)$-$(6) show R.A. and decl. in degrees, positional error in arcseconds, and off-axis angle in arcminutes. Columns (7)$-$(11) give net counts in the full band, the associated 1 $\sigma$ equivalent uncertainty based on Poisson statistics, background counts in the extraction region, net counts in the hard band, and the fraction of the PSF used for source extraction. PSF fraction values below $\sim 0.9$ indicate source crowding. Column (12) presents the $P_B$ probability of photometric significance when they lie in the $-5 \la \log P_B \la -3$ range.  

Column (13) of Table~\ref{tbl:catalog_stub} gives the sourceÕs median energy $E_{med}$ in the full spectral band after background subtraction. The median energy is a reliable surrogate for interstellar gas absorption.  For $E_{med} \ga 1.7$~keV, an empirical relationship with the gaseous column density $\log N_H = 21.22 + 0.44 E_{med}$ cm$^{-2}$ is found for pre-main sequence stars in the Orion Nebula Cluster \citep[][see their Figure 8]{Feigelson05}.  This relationship is also found in a large population of M~17 stars (Getman et al., in preparation) and we assume it applies for NGC~6334 stars here.   For normal interstellar gas-to-dust ratios, the column density can be converted to a visual absorption with the traditional relation $A_V=N_H/2 \times 10^{21}$~mag \citep{Gorenstein75} or the revised relation derived from X-ray absorption measurements, $A_V = N_H/1.6\times 10^{21}$~mag \citep{Vuong03}.   We use these calibrations between the instrumentally dependent $E_{med}$ and physical absorption measures repeatedly in the science sections below.  

\subsection{Stellar Counterparts \label{counterparts.sec}}

Unfortunately, no wide-field, sensitive, and high-resolution near-infrared image has been published covering the ACIS fields of view in NGC 6334.  We thus use the 2MASS Point Source Catalog to find stellar counterparts to the X-ray sources.  Column (14) in Table~\ref{tbl:catalog_stub} gives the $K_s$ magnitude of 2MASS stars lying within 1\arcsec\/ of the X-ray source positions.  About half of the {\it Chandra} sources have a 2MASS star within 1\arcsec\/, but the majority of these have inaccurate photometry due to nebular contamination, star crowding, or sensitivity limits.   

\begin{figure}[htbp]
\begin{center}
\includegraphics[width=3.4in]{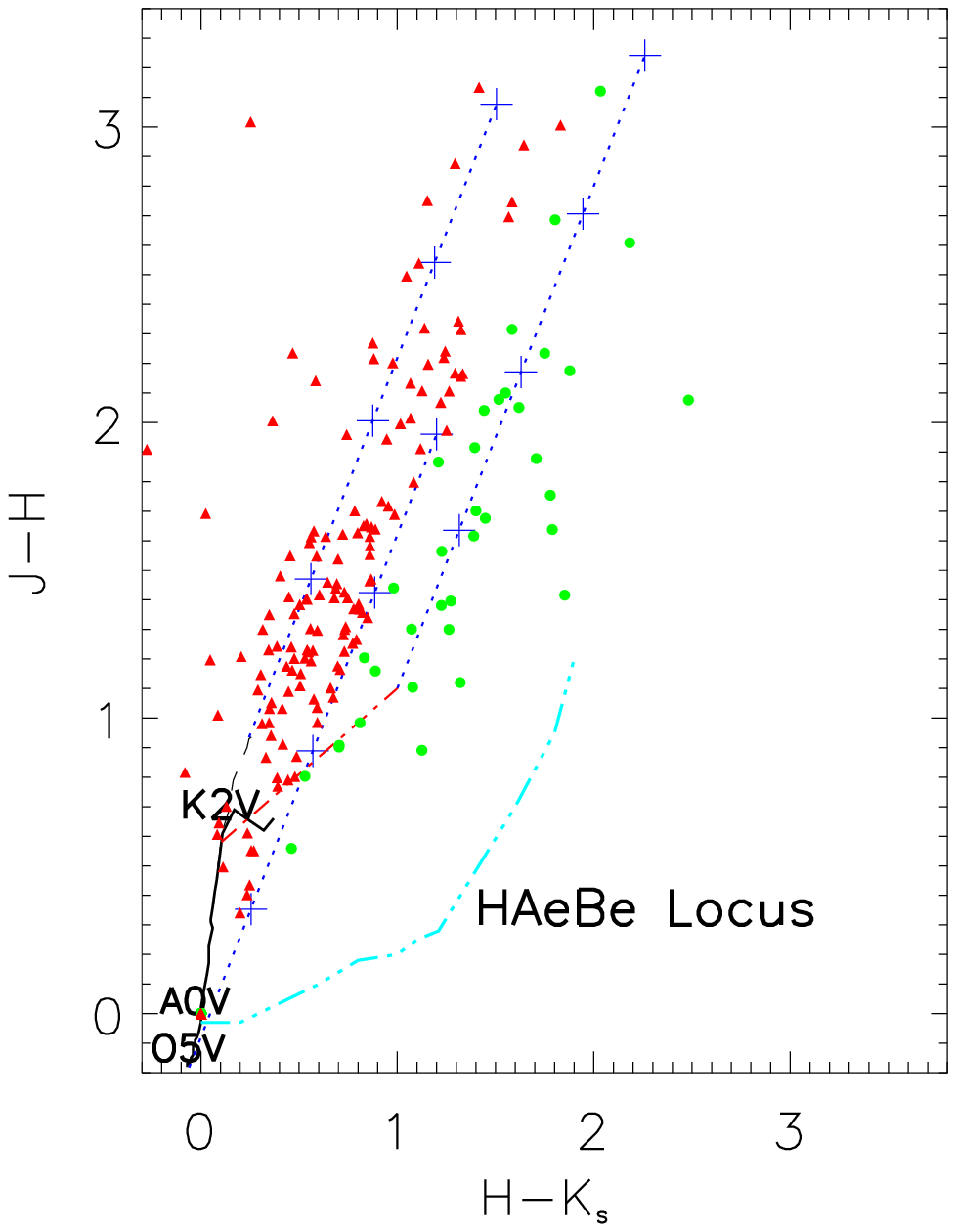}
\includegraphics[width=2.8in]{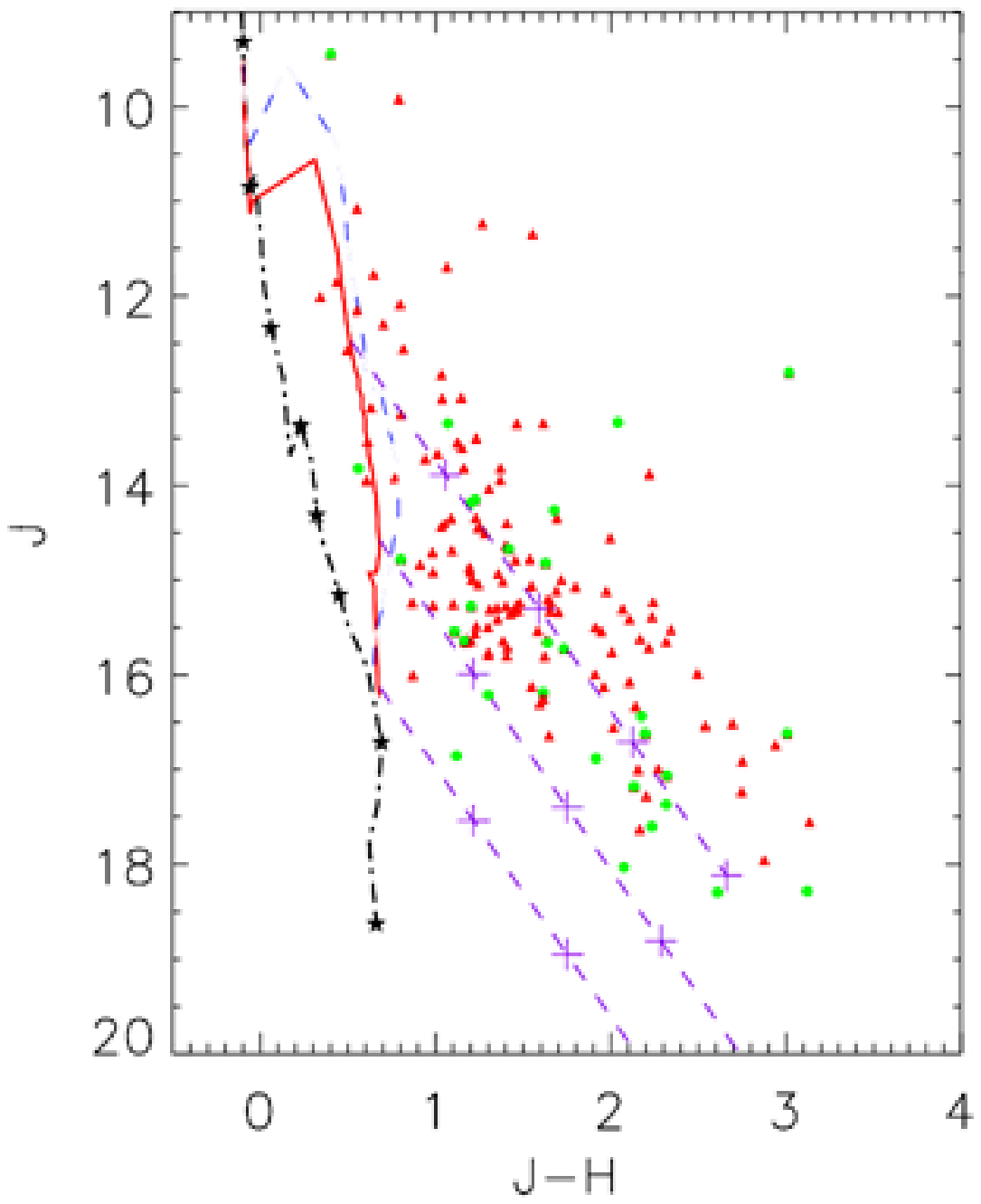}
\caption{Near-infrared properties of candidate 2MASS counterparts of the {\it Chandra} X-ray sources.
Left: Color-color diagram for counterparts having high-quality photometry. Red triangles show stars with colors consistent with reddened photospheres (Class~III stars), and green circles show stars with $K_s$-band excesses associated with protoplanetary disks (Class~II stars). Dotted lines show reddening vectors marked every $A_V = 5$~mag.  The locus of main sequence stars is from \citet{Bessell88}, classical T Tauri stars from \citet{Meyer97}, and HAeBe stars from \citet{Lada92}.  Right:  {\it J} vs. {\it J-H} color-magnitude diagram of these stars. 1~Myr and 2~Myr isochrones for pre-main sequence stars from \citet{Siess00} are shown in addition to the ZAMS.  Note that only a small fraction of the 1607 {\it Chandra} sources are shown in these diagrams 
due to limitations of the 2MASS catalog. }
\label{fig:colcol}
\end{center}
\end{figure}

Figure~\ref{fig:colcol} shows $JHK$ color-color and color-magnitude diagrams for the 2MASS stars with photometric errors better than $\pm 0.1$~mag in all three bands.  The results are similar to diagrams obtained from other {\it Chandra}-selected populations in clusters with distances around $1.5-2$~kpc such as M17 and the Rosette Nebula \citep{Broos07, Wang08, Wang09}.  The majority of X-ray selected stars do not exhibit $K_s$-band excesses attributed to the heated inner edges of protoplanetary disks; most would be classified as Class~III pre-main sequence stars based on near-infrared photometry.  The 2MASS photometric diagrams in Figure~\ref{fig:colcol} show a wide range $0 < A_V < 25$~mag.  Noting that the X-ray hardness measure $E_{med}$ exceeds 3.0 keV ($\log N_H \ga 22.5$ cm$^{-2}$ or $A_V \ga 20$ mag) for many of the X-ray sources without 2MASS counterparts, stellar counterparts for many X-ray sources will be too heavily absorbed to be present in the 2MASS catalog.  Our experience with M17 and Rosette where deeper and higher-resolution near-infrared imaging was available also indicates that the 2MASS catalog includes only a fraction of the true stellar population represented by the {\it Chandra} sources biased towards those with higher bolometric luminosities, lower obscuration, and locations away from bright nebular emission.  Due to these limitations of the 2MASS counterparts, we do not pursue scientific interpretation of these near-infrared sources (e.g., disk fractions, Initial Mass Function).  

The final column of Table~\ref{tbl:catalog_stub} gives possible associations between X-ray sources and objects from published objects in NGC~6334.  These were obtained from the SIMBAD bibliographic database using a $\pm 3$\arcsec\/ positional agreement.  The most common possible counterparts are from the shallow $JHK$ survey of \citet{Straw89} with labels FIR-I to FIR-V corresponding to the major far-infrared clusters.    We emphasize that some of these associations may not be correct due to positional inaccuracies or offsets between the X-ray and long-wavelength bands. 

\subsection{Completeness and Contaminants \label{complete.sec}}

Our source detection method is complete down to a statistical probability limit which can be converted to an ACIS count limit and, with considerable uncertainty, an X-ray flux limit.  The faintest on-axis sources in Table~\ref{tbl:catalog_stub} have $3-4$ counts.  Based on comparison with other pre-main sequence studies of pre-main sequence populations \citep[e.g., the sensitivity equation (7) in][]{Feigelson02}, this corresponds to an absorption-corrected luminosity in the $0.5-8$~keV of $L_{t,c}\sim 4 \times 10^{29}$ erg~s$^{-1}$ for absorption around $\log N_H \sim 21-22$ cm$^{-2}$.  This limit is quite approximate, as it depends on the intrinsic spectrum and cloud absorption; the limit will be a factor of $\sim 2$ lower for lightly-absorbed sources, several times higher for sources with $\log N_H \sim 23$ cm$^{-2}$, and roughly $\sim 100$ times higher for sources with $\log N_H \sim 24$ cm$^{-2}$.  Significant instrumental spatial variations in sensitivity are also present due to off-axis telescope vignetting and degraded point spread function in each field and to the overlap region between the two exposures. While it is thus difficult to establish a single X-ray completeness limit for the NGC~6334 observations for these reasons, we are confident the most of the observed region is surveyed to a sensitivity better than $1 \times 10^{30}$ erg~s$^{-1}$ except for sources suffering extremely high absorption.  

Based on the well-established correlation between X-ray luminosity and mass \citep{Telleschi07}, this sensitivity limit is sufficient to capture most of the cluster members with masses above 1~M$_\odot$ \citep[except for a minority of A and B stars,][]{Stelzer06}, and some stars between $0.3-1$~M$_\odot$.  A survey complete down to $\sim 1$~M$_\odot$ detects about 5\% of the stellar population (including young brown dwarfs) assuming a standard Initial Mass Function \citep{Chabrier03}.  We can thus multiply the number of X-ray source by a factor of $\sim 10-20$ to estimate the total population of each cluster.  This is only an approximation, as the X-ray sensitivity varies with absorption across the field and in the radial direction.  

Our method of locating {\it Chandra} sources is considerably more sensitive than that used by \citet{Ezoe06}, but it entails several sources of error.   Careful examination of the image suggests that several sources located in the wings of strong sources may be artifacts of the processing.  We cannot immediately exclude these as spurious because studies of other young stellar clusters establish that close multiple X-ray sources can appear around massive stars \citep[see HD~46150 in the Rosette's NGC~2244,][]{Wang08}.  Some of the faintest sources are also likely not real, particularly in the densest cluster regions.  Recalling that $\sim 10-20$ very X-ray faint stars should be present for each detected source, many low mass clusters members are expected to produce $1-3$ counts which are viewed as background by {\it ACIS Extract}.  In addition, truly diffuse plasma emission from shocked O star winds may present, as discussed by \citet{Ezoe06}.  The result of the elevated background is to reduce the statistical significance of real sources and to cause the values of $\log P_B$ to fluctuate with each iteration of {\it ACIS Extract} as different candidate sources are rejected and their photons enter the background.  This effect generally leads to a more conservative list of faint sources than an environment without emission from diffuse plasma or unresolved stellar populations. Nonetheless, very weak sources in the most crowded regions have an increased chance to be spurious. Altogether we believe that no more than a few percent of the sources in Table~\ref{tbl:catalog_stub} are spurious.  Improved methods for pruning the source list in crowded regions are under development (Broos et al., in preparation).  

A small fraction of the reliable X-ray sources are not members of the NGC~6334 population.  Detailed modeling of contaminant populations in the foreground or background of Galactic Plane molecular clouds at distances around $1-2$~kpc have been made by \citet{Getman06} and \citet{Wang07}.  Scaling these results to the exposures, survey area and distance of NGC~6334, we estimate that $25-50$ sources are extragalactic sources (mostly quasars), $15-30$ are foreground field stars, and $<10$ are background field stars.   Thus, $2-5$\% of the 1607 sources in Table~\ref{tbl:catalog_stub} are probably contaminants.  However, these sources must be randomly distributed across the field and their contribution to the stellar clusters discussed here will be negligible.

\subsection{Defining X-ray clusters \label{def_clus.sec}}

Cursory examination of the ACIS images in Figures~\ref{fig:ACIS}-\ref{fig:expanded} shows that the spatial distribution of X-ray sources is very inhomogeneous with distinct strong concentrations that represent young stellar clusters.  This was previously found by \citet{Ezoe06};  see their Figures~2-3.  Adopting the procedure used by \citet{Wang09}, we locate clusters on a map of the X-ray source density obtained by smoothing the locations of the 1607 sources with a Gaussian kernel of width $\sigma = 45$\arcsec\/ (0.4~pc).  Here we ignore the few sources that are likely non-members of the cluster, and consider the X-ray source surface density to represent the stellar surface density of the NGC~6334 population.  Note that these are not maps of smoothed X-ray emission; rather, the X-ray stars are viewed individual stars tracing the spatial structure of the full young stellar population.  \citet{Wang09} show that the existence of structures smaller than several arcminutes is not strongly affected by inhomogeneities in the X-ray sensitivity across the mosaic due to reduced sensitivity far off-axis of each pointing and increase sensitivity in the overlap region between adjacent images.  Nonetheless, these variations in sensitivity reduce the accuracy of our estimates of cluster richness.

\begin{figure}[htbp]
\begin{center}
\includegraphics[width=5in]{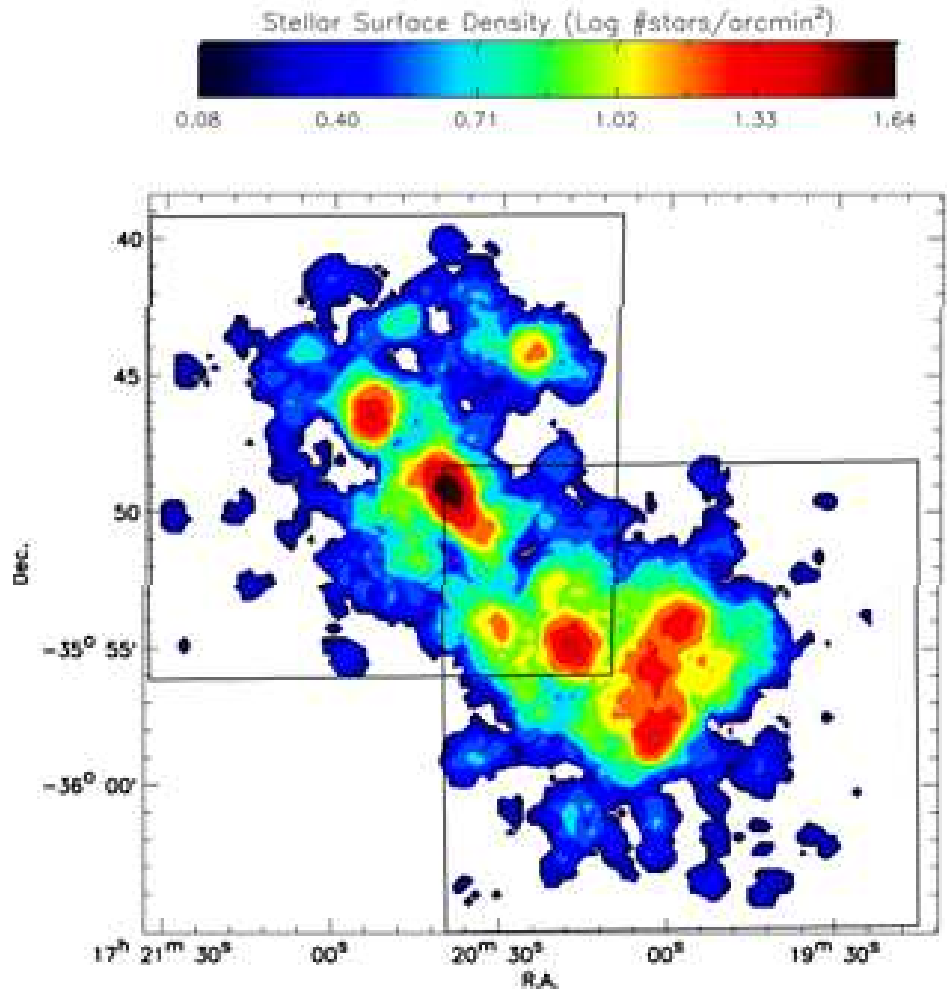}
\caption{The spatial distribution of the 1607 X-ray sources shown as a stellar surface density map after smoothing with a $\sigma=45\arcsec$\/ Gaussian kernel. }
\label{fig:space}
\end{center}
\end{figure}

\begin{figure}[htbp]
\begin{center}
\includegraphics[width=5in]{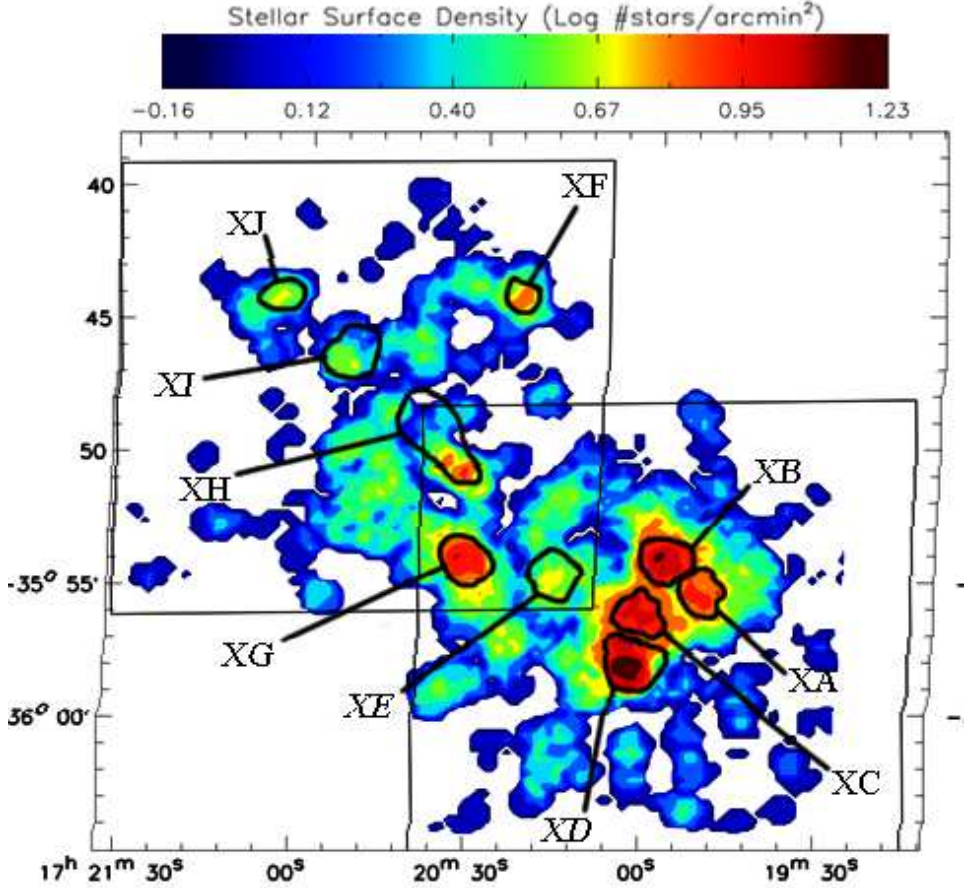} 
\end{center}
\caption{The smoothed spatial distribution of soft X-ray sources (first panel) and hard X-ray sources (second panel).  The sample is divided at $E_{med} = 2.5$ keV which approximately corresponds to column density $\log N_H \sim 22.3$ cm$^{-2}$ and visual absorption $A_V \sim 10$ mag.  The ten X-ray clusters are outlined and labeled. }
\label{fig:space2}
\end{figure}

The resulting smoothed stellar surface density map is shown in Figure \ref {fig:space} for the total $0.5-8$~keV Chandra band, and in Figure~\ref{fig:space2} for a `soft' and and `hard'.  Here we choose to divide the sample into soft sources with $E_{med} < 2.5$ keV and hard sources with $E_{med} > 2.5$ keV; the boundary approximately corresponds to column density $\log N_H \sim 22.3$ cm$^{-2}$ and visual absorption $A_V \sim 10$ mag.  Hereafter, we refer to these subsamples as the `lightly obscured' and `heavily obscured' X-ray stars and clusters.    

We have chosen to define clusters as localized peaks in X-ray source surface density which are at least 10-fold above the background level.  This corresponds to the yellow contour in Figure~\ref{fig:space} where dark blue represents the background level.   Eight or nine clusters can be discerned in this total band map.  However, examination of the smoothed soft and hard band stellar surface density maps shows dramatic differences; most clusters are prominent only in one band (Figure~\ref{fig:space2}).   From examination of the three smoothed maps, we emerge with a list of ten clusters listed in Table~\ref{tbl:xray} and labeled in Figure~\ref{fig:space2}.   The table columns give our designations XA through XJ, central location (accurate to $\pm 20\arcsec$), indicator of X-ray hardness, approximate extent based on the yellow contour in Figure~\ref{fig:space}, and the number of X-ray sources found inside this contour for each X-ray cluster.  Recall that 1\arcmin\/  corresponds to about 0.5~pc at a distance of 1.7~kpc.

The procedure used here to define X-ray selected clusters based on peaks in a smoothed distribution is not unique.  Other statistical techniques can be used: smoothing using different kernel widths; nearest-neighborhood methods \citep{Roman08}, single-linkage agglomerative clustering (also called the `friends-of-friends' algorithm),  maximum-likelihood mixture models assuming Gaussian cluster shapes, and so forth.   Each of these methods has advantages and disadvantages, and none gives a unique `correct' result.  \citet{Wang09} compares our procedure to a 10-th nearest neighbor algorithm in a {\it Chandra} mosaic of the Rosette Molecular Cloud, and shows that the results are very similar.  For any clustering method, the results may not be physically meaningful.  For example, it is possible that the four clusters XA-XD represent a single physical structure that appears divided by inhomogeneous obscuration.  The confusing effects of variable extinction in NGC~6334 was raised by \citet{Lindsay55} and others. Our trials in constructing the cluster list with different smoothing and associations between hard and soft structures gave $8-12$ clusters.   We conclude that the division of the X-ray sources into ten distinct clusters is a reasonable, but far from unique, result.

\subsection{Comparison to Long-Wavelength Studies of NGC 6334 \label{long.sec}}

Figure~\ref{fig:overlay} superposes the smoothed X-ray source distributions on wide-field maps at radio and far-infrared wavelengths.  The top panels show contours from the Very Large Array 18~cm continuum map of \citet{Sarma00} with labels for the NGC~6334~A-F HII regions identified by  \citet{Rodriguez82}.  Higher resolution radio maps of some components are presented by \citet{Carral02}.  The bottom panels show a map at 71~$\mu$m by \citet{Loughran86} obtained with a balloon-borne telescope with their designations of far-infrared sources I-V.  Somewhat higher resolution far-infrared maps from HIRES analysis of the {\it IRAS} satellite data appear in \citet{Kraemer99}.  

Examination of these map overlays shows that there is a poor correspondence between lightly obscured X-ray clusters and radio/infrared structures, but a good correspondence when heavily obscured X-ray clusters are considered.  Thus, XE appears associated with far-infrared IV and radio A, XH is associated with II and D, and XI is associated with far-infrared I and radio E and F.  The lightly obscured clusters XA-XD, XF, XG, and XJ have little association with long wavelength structures, although some of the stars we include with XC-XD appear embedded and may be associated with far-infrared V and the southwestern radio shell.  
 
Figure~\ref{fig:GLIMPSE} shows the X-ray clusters superposed on the mid-infrared emission observed by the Spitzer GLIMPSE survey \citep{Benjamin03}.  Some association can be seen with the heavily obscured clusters:  XE lies in a confused region of bright and dark infrared nebulosities; XH is straddled by large bubbles emerging from the ridge line; and XI coincides with an isolated bright region within a larger dark cloud.  There is also a hint that the massive stars in lightly obscured X-ray clusters (particularly XA, XC, XD and XG) have produced bowl-like cavities in the cloud material traced by the Spitzer map.  But the relationship between the X-ray maps (which mostly represent lower mass stars) and the mid-infrared maps (which mostly represent heated dust and PAH molecules) is indirect.  We do not significantly use the GLIMPSE map in the science analysis of this study, as we focus on the stars rather than the interstellar material.  

We proceed with a more detailed examination of individual stars and star groups in various components of NGC~6334, proceeding west to east. 

\section{STARS AND STELLAR CLUSTERS IN THE CHANDRA POPULATION} \label{sec:xcluster}

\subsection{Embedded Stars in NGC~6334~V \label{Vregion.sec}}

A group of faint embedded X-ray sources around ($\alpha,\delta$)=($17^h19^m58^s, -35\arcdeg58\arcmin$) lying behind the lightly absorbed cluster XD can be seen in the hard band image.   Its surface density was not high enough to be classified as an X-ray cluster.  The group includes sources \#218, 219, 226, 228, 234, 236, 248, 249, 252, 266, 270, 282 and 294.  Several of the sources exhibit extraordinarily high X-ray absorptions with $4 < E_{med} < 7$ keV, corresponding to $23.0 \la \log N_H \la 24.3$ cm$^{-2}$ and $50 \la A_V \la 1000$ mag based on the calibration discussed in \S\ref{srcID.sec}.  In nearby star formation regions, only a handful of young stellar objects with dense localized envelopes have comparable X-ray absorptions including the Class 0/I sources CrA IRS~7 and IC1396N BIMA~2 \citep{Hamaguchi05, Getman07}.  

These X-ray sources lie on the southeast side of the NGC~6334~V far-infrared region and surround the embedded B-type protostars producing masers and the infrared emission nebulae IRN~V-1 and IRN~V-2 \citet{Hashimoto08}.  These massive protostars are too heavily obscured to be seen in the near-infrared, consistent with the extreme absorption detected in the nearby X-ray sources which probably represent the most X-ray luminous low mass members of the associated cluster.  This grouping bears some resemblance to the cluster of $\sim 20$ X-ray emitting low mass stars associated with the cluster of B-type protostars in the Kleinman-Low Nebula behind the Orion Nebula with $\log N_H \simeq 23.0$ cm$^{-2}$ \citep{Grosso05}.   

A sparse but compact (diameter $< 0.1$ pc) group of faint, moderately absorbed X-ray sources appears to be associated with the bright H$\alpha$ emission line star SS~303 (K=8.57) and the infrared source NGC 6334 V IRS 1.  These include \# 302, 305, 306, 310, 311 (= SS~303) and 320 (= NGC~6334~V IRS1). \citet{Jackson99} suggest that SS~303 ionizes the the southeastern radio shell, G351.20+0.70.  Another possible ionizing source is the bright star \#210 = FIR-V 51 (K=7.76), although it lies off-center within the radio shell.

\subsection{Lightly Obscured Clusters XA-XD \label{XA-XD.sec}}

The XA, XB, XC and XD clusters together constitute the largest and richest complex seen in the NGC 6334 region in the X-ray band.  It is unclear whether they are structurally distinct structures or part of a single larger cluster subject to inhomogeneous absorption.  The latter interpretation is supported by the smoother appearance of the source distribution in the hard band than in the soft band (Figure~\ref{fig:space2}).  These X-ray sources are lightly absorbed with E $\le$ 2.5 keV, and relatively few are present in the hard X-ray maps.  This obscuration corresponds to $\log N_H \la 22.3$ cm$^{-2}$ and $A_V \la 10$ mag.  

These X-ray structures have few published counterparts in other wavelengths, perhaps because they are older clusters and no longer embedded in dense cloud cores.   The mid-infrared source IRAS 17175-3554 corresponds closely with XC, although we suspect is lies behind the cluster in the cloud interior.  Cluster XD can be seen in the near-infrared H$_2$ and $L$ band images obtained by \citet[][see their Figure 5]{Burton00} where it was associated (perhaps incorrectly) with the embedded far-infrared source NGC~6334-V.   Many of the X-ray sources with 2MASS counterparts with $11 < K < 14$ are likely low mass cluster members, and some are FIR-V stars from \citet{Straw89}.   A number of stars bright in both the X-ray and $K_s$ bands are likely high mass cluster members.  These include \# 210 (= FIR-V 51, see \S\ref{Vregion.sec}), 354 (=FIR-V 10), 405 (= FIR-V 102), 415 (= FIR-V 107), 505, 564, and 586.   Source \#405 = FIR-V 102 ($K_s = 8.81$) lies in a compact group (diameter $\sim 0.2$~pc) of $\sim 20$ X-ray sources around ($17^h20^m03^s$,$-35\arcdeg 58\arcmin20\arcsec$) that can be clearly seen in the X-ray image shown in Figure~\ref{fig:expanded} (bottom panel).  This group at the center of the XD cluster has the highest surface density in the NGC 6334 complex.

\subsection{Heavily Obscured Cluster XE (HII region NGC~6334~A; FIR NGC~6334 IV) \label{XE.sec}}

XE is the most prominent cluster seen in hard X-rays in the southwest quadrant of the NGC~6334 complex.  It is only very weakly detected in soft x-rays. XE is spatially associated with the radio HII region NGC~6334 A and the far-infrared stellar cluster NGC~6334 IV.  This positional agreement confirms that the hard X-ray band can readily detect and resolve heavily embedded clusters as seen, for example, in Cepheus~B \citep{Getman06}, W3~Main \citep{Feigelson08}, and the Rosette Molecular cloud \citep{Wang09}.

XE is a rich and concentrated cluster of obscured X-ray sources associated with far-infrared source FIR~IV and the bipolar radio source A.  X-ray $E_{med}$ ranges from around 2 to an extraordinary $E_{med} \sim 5-6$ keV, corresponding to $\log N_H \sim 22.3$ to 23.8 cm$^{-2}$ or $A_V \sim 10$ to 300 mag. These unusually heavily obscured sources in XE include \#607, 669 (= FIR-IV 43), 692 (= FIR-IV 60), 712 and 715.  They likely lie within the dense, $\sim 1$~pc-diameter molecular clump emitting brightly in high-excitation CS rotational emission lines \citep{Kraemer99}.    As with the group around IRN~V-1(\S\ref{Vregion.sec}), we emphasize that young stars are rarely seen with $E_{med} \ga 5$ keV in other star forming regions \citep[see, for example, the Chandra Orion Ultradeep Project $E_{med}$ distribution in][]{Feigelson05}.  

Several of these heavily absorbed sources lie near the millimeter sources identified by \citet{Sandell99}.  Sources  \#737 and 739 lie near MM~1a, \#712 lies near MM~1b and MM~2,  and \#712 lies near MM~2.  Source \#703 with $E_{med}=3.0$ keV (corresponding to $\log N_H \sim 22.5$ cm$^{-2}$ and $A_V \sim 20$) coincides with the infrared star IRS~20 associated with infrared nebula IRN~IV~2 \citep{Hashimoto08}. The region around MM-3 is studied by \citet{Persi09}; only one of their embedded infrared-excess stars appears as an X-ray source (\#757 = IRS 9).   There is no concentration of sources around the $\sim 20$\arcsec\/ shell-like radio HII region NGC~6334~A discussed by \citet{Carral02}. 

The hard band X-ray cluster XE has a very high stellar surface density and and inferred total stellar  population of $> 500-1000$ stars when undetected stars are considered (\S\ref{complete.sec}).  The GLIMPSE image of the region shows a complex of irregular dense dark clouds superposed on bright emission nebulosity.  The elongated obscuring material resemble Infrared Dark Clouds (IRDCs) noted in other star forming regions \citep[e.g.][]{Carey98, Rathborne06}.   Some of the heavy absorption seen in X-ray stars may be due to these dense obscuring irregular clouds.  

\subsection{Extragalactic Blazar RCM B \label{RCMB.sec}}

The very bright X-ray source \#787 coincides with the compact radio continuum source NGC 6334 B (= RCM B = IGR J17204-3554) which has been widely discussed as a radio-loud active galactic nucleus lying behing the NGC~6334 star forming region \citep{Rodriguez88, Trotter98, Bassani05,  Bykov06}.  This source is an order of magnitude too bright for even the most extreme flaring pre-main sequence stars, and has no stellar counterpart corresponding to a cluster OB star.  Its ACIS median energy $E_{med} = 4.7$~keV (Table 1) indicates heavy absorption corresponding to $log N_H = 23.3$~cm$^{-2}$ ($A_V \sim 100$~mag) consistent with the source lying behind the NGC~6334 cloud and other Galactic Plane material.

\subsection{Lightly Obscured Cluster XF \label{XF.sec}}

NGC~6334 XF is a small and sparse cluster $\sim$0.5~pc across  seen best in soft X-rays.  It lies 2~pc (projected) north of the star forming ridge that defines the NGC~6334 complex.  Cluster membership appears to be dominated by source \#821 ($K_s=5.71$), \#846 ($K_s=7.36$), and \#881 (= CD~-35$^\circ$11482 = IRAS 17170-3539, $K_s=7.76$, B0.5e).  Source  \#881, with $>$500 ACIS counts, is the brightest X-ray source in the NGC~6634 complex after the O7 star \#972 = FIR-III 13 in cluster XH (\S\ref{XH.sec}).  The unusually bolometrically bright stellar counterpart of \# 821 may be a supergiant like HD~319701 and HD~319702 (\S\ref{unclustered.sec}).  If correct, this suggests that the XF cluster is older than most of the other clusters with its most massive member evolved off of the main sequence.  

X-ray absorptions of XF members are typically in the range $E_{med} \simeq 1.5-3$ keV corresponding to $\log N_H \sim 21.9-22.5$ cm$^{-2}$ and $A_V \sim 5-20$ mag.  The Spitzer GLIMPSE image of this region (Figure~\ref{fig:GLIMPSE}) shows very dark absorption bands characteristic of IRDCs.   The GLIMPSE image shows bright nebular emission throughout the vicinity (previously labeled NGC 6334~I(NW), see Figure~\ref{fig:overlay} bottom left panel), perhaps with an IRDC superposed.

\subsection{Lightly Obscured Cluster XG \label{XG.sec}}

This cluster in similar to XF, lying off of the molecular ridge and appearing only in the soft X-ray image. The cluster is somewhat richer than XF with $\sim 25$ X-ray sources, implying a total population around $250-500$ members.   It is the only x-ray cluster that is located to the south of the star forming ridge of NGC~6334.   Source \#911 is an unusually bright X-ray emitter in the cluster.  Its stellar counterpart is a heavily reddened star with $R-K=6.6$ mag from DENIS catalog photometry.  Cluster members have a range of absorptions centered around $E_{med} \sim 2$~keV, corresponding to $\log N_H \sim 22.0$ cm$^{-2}$ and $A_V \sim 5$.

\subsection{Heavily Obscured Cluster XH  (HII region NGC~6334 D; FIR NGC~6334 II)
\label{XH.sec}}

XH has the highest stellar surface density in the hard X-ray image.  It is
moderately heavily absorbed with typical $E_{med} \sim 2-3$ keV corresponding to 
$\log N_H \sim 22.0-22.5$ cm$^{-2}$ or $A_V \sim 5-20$ mag.  A few of its sources
appear in the soft X-ray band on the southwest side; this might be a separate sparse
cluster.  The X-ray concentration accurately coincides with the far-infrared source 
NGC~6334~II and, displaced $\sim 1$\arcmin\/ east of the X-ray peak, with the large 
radio HII region NGC~6334~D (Figure~\ref{fig:overlay}).  This HII region appears as a 
bubble-like structure in the GLIMPSE map (Figure~\ref{fig:GLIMPSE}) and other tracers 
of heated interstellar material.

The brightest star in XH seen in both the near-infrared and X-ray bands is
source \#972 coincident with the $K_s=7.39$ star known variously as TICID 7374-463 
from the Tycho Input Catalog, FIR-III~13 from the near-infrared photometric
survey of \citet{Straw89}, and IRN~III-1 from the near-infrared polarimetric
survey of \citep{Hashimoto08}.  With 1653 ACIS counts, it is the brightest X-ray source 
in NGC~6334, appearing several arcminutes off-axis of both ACIS fields.
It was marginally detected as the {\it Einstein Observatory} source 2E~1717.1-3548.
As it lies 2\arcmin\/ southwest of the dense core of XH and is only moderately
absorbed with $E_{med}=1.8$~keV, this star may be better interpreted as a
isolated O star rather than as a member of XH.  It may be the principal ionizing
source of the northern portion of the NGC~6334~C HII region.

Other bright stellar counterparts to X-ray sources in XH include \#1291 (= FIR-II-23) 
and 1312 with $K_s \sim 9$, and  \#953 (= FIR-III~10), 1051, 1107, and 1275 
(= FIR~III~24) with $K_s \sim 10-11$.

\subsection{Heavily Obscured Cluster XI  (HII regions NGC~6334 E and F; \\ FIR NGC~6334 I)
\label{XI.sec}}

NGC~6334 XI is a small, sparse cluster seen in the hard X-ray band along the 
molecular ridge of NGC~6334.   It corresponds closely to the far-infrared
source I (= IRAS~17175-3544) and is centered between the two radio
continuum HII regions NGC~6334~E and NGC~6334~F (Figure~\ref{fig:overlay}).  
The HII region NGC~6334~E, centered an arcminute north of the X-ray cluster, 
has a nearly circular radio continuum shell with $\sim 30$\arcsec\/ diameter
and a compact radio source at the middle associated with a very heavily
absorbed near-infrared star TPR 161 \citep{Carral02}.  This massive star does 
not appear in the {\it Chandra} image and there is no apparent concentration 
of lower mass stars within the radio HII region.

Although the XI cluster appears centrally concentrated in the smoothed
maps of Figures~\ref{fig:space}-\ref{fig:space2}, examination of the
individual source positions in Figure~\ref{fig:expanded} suggests the
cluster may be bifurcated with subclusters centered at  ($17^h20^m52^s$, $-35\arcdeg45\arcmin50\arcdeg$) and ($17^h20^m53^s$, $-35\arcdeg46\arcmin20\arcsec$).  The southern X-ray subcluster in cluster XI lies directly on the unusually strong far-infrared source NGC~6334~I with a 60~$\mu$m flux density $\sim 11,000$~Jy and a bolometric luminosity $\sim 2 \times 10^5$~L$_\odot$ 
\citep{Sandell00}.  This is the far-infrared source NGC~6334~I  and
IRAS source 17175-3544.  A cluster of near-infrared stars extends over 2\arcmin\/
\citep{Tapia96}.  

At millimeter wavelengths, the core of NGC~6334~I is resolved 
into several sources, some associated with a cometary ultracompact HII
region in radio continuum and masers \citep{Hunter06}.  This is clearly a
concentrated subregion of current massive star formation.  X-ray sources \# 1428, 
1429, 1430, 1431, and 1435 lie within a $\sim 3$\arcsec\/ region associated
with NGC~6334~I-SMA3, IRS-1E, IRS-I~1 and the bright head of the radio
continuum structure. See \citet{Hunter06}, \citet{Beuther07}, \citet{Beuther08}, 
and \citet{Seifahrt08} for  detailed multiwavelength maps of this crowded subregion.  
Two X-ray sources in this crowded region have counterparts within 1\arcsec\/
at long wavelengths.  Source \#1428 with $E_{med}=2.8$~keV (corresponding 
to $\log N_H \sim 22.5$ cm$^{-2}$ or $A_V \sim 20$ mag) coincides with the 
star IRS-I-1 \citep[Keck II OSCIR designation,][]{deBuizer02} or IRS1-W
\citep[Very Large Telescope NACO designation,][]{Seifahrt08} with estimated spectral type B2.  
Source \#1429 with $E_{med} \simeq 3.0$~keV coincides with IRS1-SW seen
with the NACO camera \citep{Seifahrt08}.  Several other X-ray
sources are present close to this dense core, and the locally high density of photons 
assigned to detector background strongly suggests that additional fainter sources 
would be resolved in a longer X-ray exposure.  

The cometary ultracompact radio HII region NGC 6334F \citep[= G351.42+0.64;][]
{Carral02} lies $\sim 10$\arcsec\/ west of the peak concentration of NGC 6334~I 
and is associated with the far-infrared source IRAS 17175-3544.   
The moderately strong and absorbed X-ray source \#1406 lies within the 
$\sim 6$\arcsec\/ radio nebula and possibly associated with the near-infrared 
star TPR~43. It could be an X-ray detection of the exciting star of NGC~6334~F.  
About 10 unassigned X-ray photons lie within a few arcseconds of \#1406, representing 
either the shocked stellar wind or emission from unresolved faint low-mass stars.  

On the northern edge of the X-ray cluster XI lies the important protostellar region called
NGC~6334~I(N).  This is a strong extended (sub)millimeter continuum emitter
associated with $M \sim 2000$~M$_\odot$ of dense molecular gas, molecular
outflow and maser emission, but no radio or infrared counterparts 
\citep[][and references therein]{Sandell00}.   It has been resolved into
a compact cluster of millimeter massive protostars \citep{Hunter06}. See 
\citet{Beuther08} for detailed multiwavelength maps of this subregion. 
Three faint X-ray sources are detected in the $\sim 30$\arcsec\/ NGC~6334~I(N) subregion:
\#1447, 1454 and 1463.  Sources \#1454 and 1463 exhibit  extraordinarily high 
X-ray absorption with $E_{med} \sim 5.5$~keV, corresponding to $\log N_H \sim 
23.5$ cm$^{-2}$ and $A_V \sim 200$ mag.  These are very likely members of the 
embedded cluster.   Source \#1454 lies within 1\arcsec of the
millimeter source SMA~7 and a faint {\it Spitzer} IRAC star, while \#1463
may be associated with the near-infrared star TPR~216 and/or methanol
masers.  Source \#1447 appears coincident with SMA~6 but, due to its
lower X-ray absorption ($E_{med} \sim 1$~keV), may instead be a foreground star.

\subsection{Lightly Absorbed Cluster XJ \label{XJ.sec}}

NGC~6334 XJ is a sparse and small cluster  ($\sim$ 1pc across)  seen only in
soft x-rays.  Its dominant star is the counterpart of \#1528, 2UCAC 16657515 
with $V=14.2$ and $K_s=9.3$.

\subsection{Absent Clusters \label{absent.sec}}
  
The long-established far-infrared source NGC~6334~III at the center of
NGC~6334 is thought to be more evolved than sources at the southwest
and northeast extremeties.  It has no masers or outflows but is associated 
with the large, diffuse bipolar HII region NGC~6334~C.  The infrared source
lies between the lightly-obscured X-ray cluster XG and the heavily obscured
cluster XH, but there is no concentration of X-ray sources clearly associated with it
or the NGC~6334~C HII region.  We suggest that no stellar cluster is
present here.

\subsection{Unclustered massive stars \label{unclustered.sec}}

A number of X-ray sources with $8 \la K \la 11$ counterparts lie 
outside the main cluster concentrations.  Several bolometrically luminous stars appear as X-ray sources west of the XA-XD clusters.  The faint X-ray source \#2 coincides with HD 319701 (K=6.76), an isolated and lightly obscured B1Ib supergiant \citep{Neckel78} lying $\sim 10\arcmin$\/ southwest of the XA-XD complex.  The faint source \#63 similarly lies southwest of the main clusters and appears to be associated with a heavily absorbed massive star.  Several other sources bright both in the X-ray and $K_s$ bands are distributed in this region are likely previously unnoticed massive stars (e.g., \# 77, 79, 108 and 130).

The bolometrically brightest of these stars are counterparts of \#564 with a $K_s \sim 8.1$ counterpart lying $\sim 2\arcmin$\/ northeast of XB, \#949 with a $K_s \sim 8.6$ counterparts lying $\sim 2\arcmin$\/ north of XG, and \#972 near cluster XH discussed in \S\ref{XH.sec}.  Other apparently massive stars distributed across the field include \# 628, 719, 726, 728, 748, 757, 817, 890, 953 and 1051.  These may be massive or intermediate-mass stars that have drifted or been ejected from one of the rich clusters.  

Source \#1371 coincides with the X-ray and bolometrically bright star HD 319702 ($K_s \sim 7.39$) on the eastern side of the NGC 6334 complex, several arcminutes south of cluster XI.  It has been spectroscopically classified as a B1Ib supergiant \citep{Neckel78} and a O8 III((f)) giant \citep{Walborn82}.  Although HD 319701 and HD 319702 were identified as the exciting stars of the NGC 6334 HII region in early work \citep[e.g.,][]{Sharpless59}, it now appears likely that they are older stars that have drifted from the main concentrations of heated molecular gas.  Other optically bright O stars commonly associated with NGC~6334 (HD 156738, 319699 and 319703) lie outside the {\it Chandra} fields of view.  HD 319703 is a bright X-ray source detected with the {\it Einstein Observatory} \citep{Chlebowski89}.

\section{DISCUSSION AND CONCLUSIONS \label{discuss.sec}}

This {\it Chandra}-based study of the stellar population in the NGC~6334 star 
forming region must be viewed as preliminary in several respects.  First, the
construction of the sensitive catalog of X-ray sources has deficiencies.  
Some of the faintest sources may be spurious, source 
properties such as X-ray luminosities are not estimated, and inhomogeneities
in the survey coverage are not quantitatively treated.  Second, the exposures
are short so that only $5-10$\% of the full Initial Mass Function is detected.  
Third,  the interpretation
of X-ray photons which are not assigned to resolved sources is not addressed
here: do they arise from fainter X-ray stars which must be present in the 
clusters (\S\ref{complete.sec}), or from diffuse plasma from shocked O star winds
\citep{Ezoe06}?  Both the incompleteness of the X-ray catalog to faint cluster
members, and the interpretation of these unassigned photons, require
substantially longer {\it Chandra} exposures than available for this study.   
Fourth, the association of X-ray sources with near-infrared
sources is inadequate.  The 2MASS catalog has insufficient resolution
to resolve crowded cluster regions and to obtain reliable photometry
in the presence of spatially complex nebular emission.  Scientific
inferences from infrared color-magnitude and color-color diagrams,
such as Initial Mass Functions and protoplanetary disk fractions, 
can not be reliably obtained.  

Despite these deficiencies, our survey represents the deepest 
penetration into the Initial Mass Function of the young population over
the full NGC~6334 region, and is effective in identifying both dozens of
luminous OB stars (many of which were not previously identified) and 
hundreds of lower mass members of concentrated stellar clusters.  These
advantages accrue in large part to the low contamination of the X-ray
selected population by extraneous sources.  The principal results are: 
\begin{enumerate}

\item A catalog of 1607 faint X-ray sources is presented, most with
subarcsecond positions.  Although only half have 2MASS counterparts, we 
infer from past studies that $\sim 95$\% of these sources are likely
cluster members.  The remaining sources are extragalactic and stars 
unrelated to NGC~6334.  The X-ray survey suffers far less 
contamination from Galactic field stars or diffuse nebular emission 
than infrared surveys, and should be nearly unbiased with respect to
protoplanetary disk emission.   In addition, we obtain estimates of 
line-of-sight absorption to each source individually from the median 
energy of its detected photons. 

\item The X-ray exposures are roughly complete to $L_x \sim
1 \times 10^{30}$ erg~s$^{-1}$. Based on X-ray studies of the stellar 
populations in the nearby Taurus and Orion clouds, this sensitivity 
captures the top $\sim 5-10$\% of the pre-main sequence X-ray luminosity 
function and most of the cluster members with masses above 1~M$_\odot$.
The total inferred population of NGC~6334 is around $20-30,000$ stars.  

\item The spatial distribution of X-ray sources is very inhomogeneous 
and differs greatly when lightly obscured sources are compared to 
heavily obscured sources.  From smoothed maps of the source 
distribution, ten X-ray selected clusters are identified, labeled XA to 
XJ.  Seven of the clusters are mostly seen in the soft X-ray band
and, with out exception, were previously unnoticed in visible, infrared 
or radio studies.  The three hard-band X-ray clusters, plus a small 
group behind clusters XC-XD, are closely associated with well-studied 
luminous far-infrared clusters.   HII regions are loosely associated with
these clusters, but are not generally concentrated in the center of
the clusters.

\item The X-ray population includes some of the previously identified
OB stars, and detected a considerable number of additional stars with
$K_s$ band magnitudes indicative of new OB members.  Some of the massive
stars are dispersed across the X-ray fields, several parsecs from the 
concentrations of current star formation.  At least two appear to be
(super)giants.  This suggests that these massive stars have either 
drifted or been ejected from their natal sites, and that star formation
has been ongoing in NGC~6334 for several million years or longer. 

\item The X-ray
population also includes some extraordinarily obscured sources with
X-ray absorption up to  $\log N_H \simeq 24.0$~cm$^{-2}$ or $A_V 
\sim 500$ mag.  It is likely that such high absorptions cannot be produced
by interstellar cloud material, but more likely by localized infalling
envelopes around protostars \citep{Getman06}. Several of these sources 
emit mostly in the submillimeter band, an indicator of Class~0 classification.
These include well-studied systems in NGC~6334~I, NGC~6334~I(N), 
NGC~6334~IV, and NGC~6334~V. If some of these X-ray sources
are truly massive Class~0 protostars known from submillimeter studies; 
they may be the youngest stars ever detected in the X-ray band.  
The only comparably embedded systems reported in the X-ray band are
the lower-mass Class~0-I sources IRS~7 in Coronal Australis cloud with $\log N_H
\simeq 23.5$~cm$^{-2}$ and IRAS~21391+5802 with $\log N_H \simeq
24.0$~cm$^{-2}$ in the triggered cloudlet IC~1396N \citep{Hamaguchi05, 
Getman07}.  We note, however, that it is difficult with the available data
to distinguish emission from the submillimeter source itself from closely
associated lower mass pre-main sequence stars. For example, the
Becklin-Neugebauer Object in the OMC-1 core behind the Orion Nebula
is a very faint X-ray source lying only $\sim 1$\arcsec\/ ($\sim 400$~AU)
from a much more X-ray luminous lower mass companion \citep{Grosso05}. 

\end{enumerate}

It is clear that there is great promise in further study of the stellar population
of NGC~6334 based the results obtained here.  
First, the existing 40~ks {\it Chandra} exposures are photon-starved, with most
detections having $<$10 counts and the vast majority of lower mass stars
undetected.  A $\ga$10-fold increase in exposure would alleviate this problem
and provide an empirical basis for interpreting the unresolved X-ray emission
as stellar or diffuse.  Second, wide-field, near- and mid-infrared images with
sub-arcsecond resolution are urgently needed for comparison with the X-ray 
source catalog.  The challenge here is less reaching high sensitivity, as the 
counterparts to the X-ray sources will mostly have $K_s \la 16$, as resolving 
stars from the bright, spatially complex nebular emission.   Third, near-infrared
spectroscopy of X-ray selected stars would provide spectral types and direct
estimates of stellar masses.  The highest priority for spectra are the $8 < K < 11$
counterparts which are likely new OB members of the cluster.

\acknowledgements  We thank Leisa Townsley (Penn State) for many
stimulating discussions.  Konstantin Getman (Penn State) and Junfeng
Wang (Harvard-Smithsonian Center for Astrophysics) gave very helpful
advice and technical assistance, and an anonymous referee provided useful
improvements.   This  work was supported by the Chandra 
ACIS Team (G. Garmire, PI ) through NASA contract NAS8-38252. This 
publication makes use of data products from the Two Micron All Sky Survey (a
joint project of the University of Massachusetts and the Infrared
Processing and Analysis Center/California Institute of Technology,
funded by NASA and NSF), the Spitzer
Space Telescope (operated by the Jet Propulsion Laboratory,
California Institute of Technology, under a contract with
NASA), and the SIMBAD database (operated by the 
Centre de Donn\'ees astronomiques de Strasbourg).


\newpage

\begin{figure}
\begin{center}
\includegraphics[width=6in]{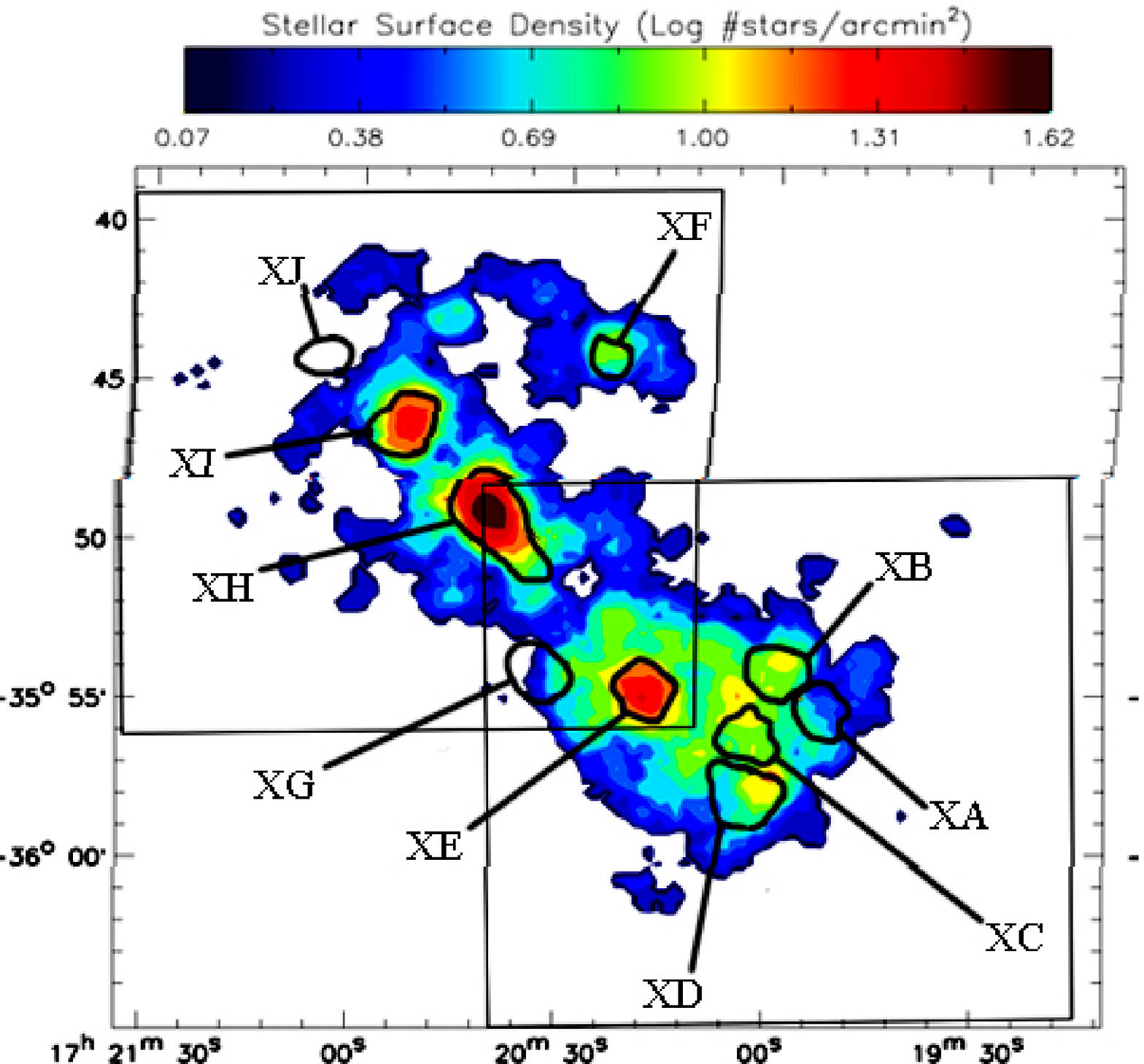}
\end{center}
\end{figure}

\begin{figure}[htbp]
\begin{center}
\includegraphics[width=3in]{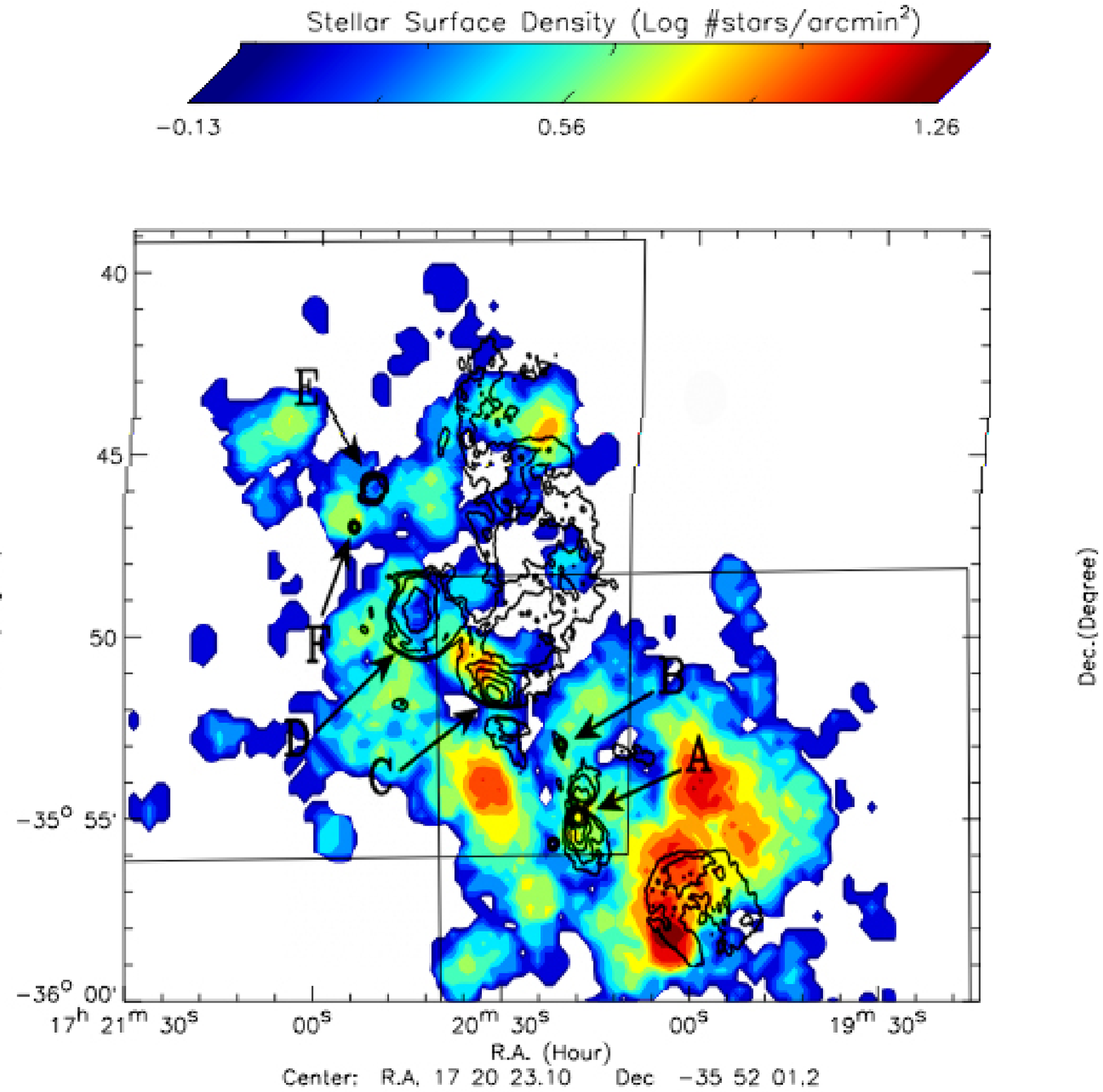}
\includegraphics[width=3in]{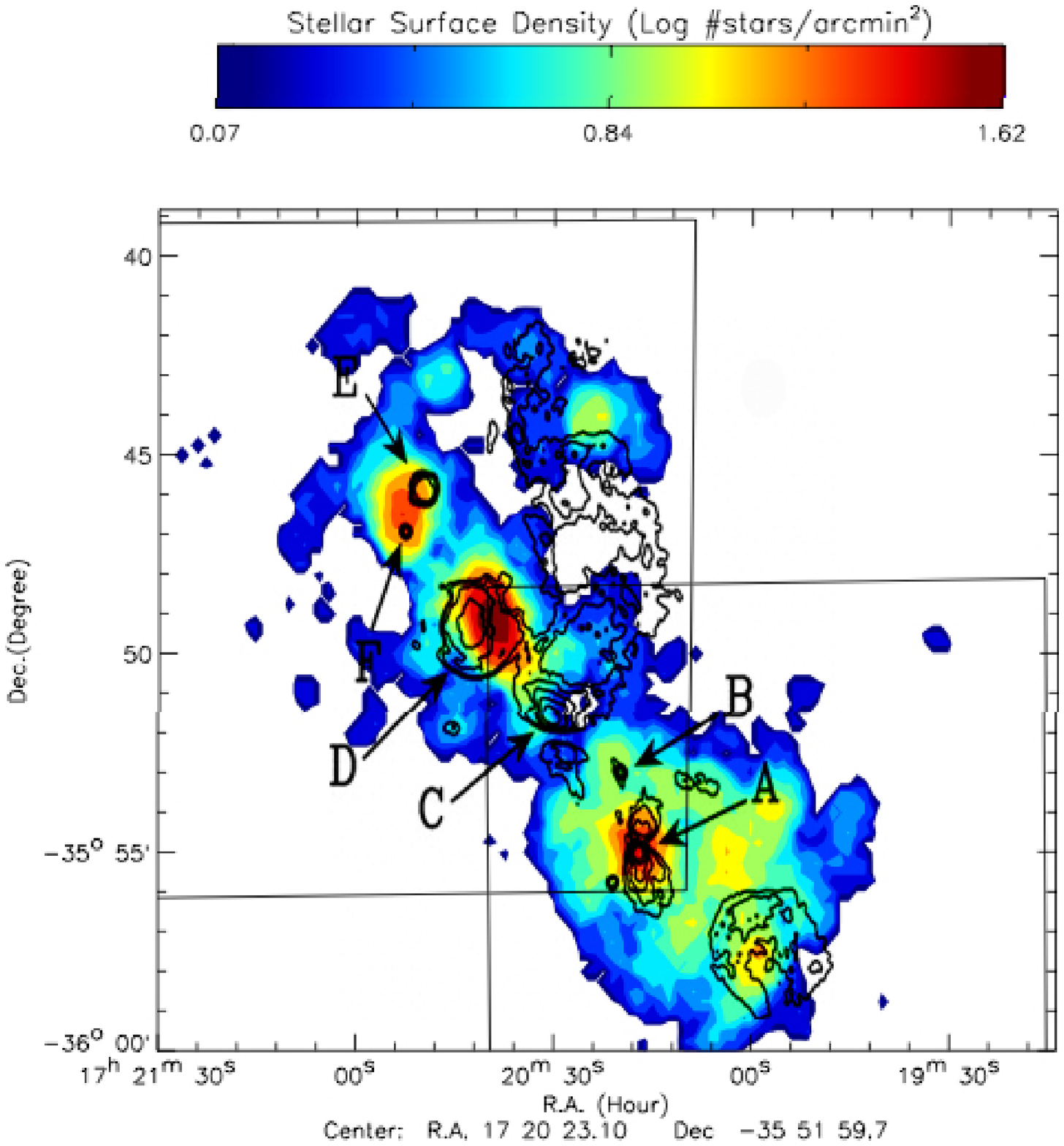}
\includegraphics[width=3in]{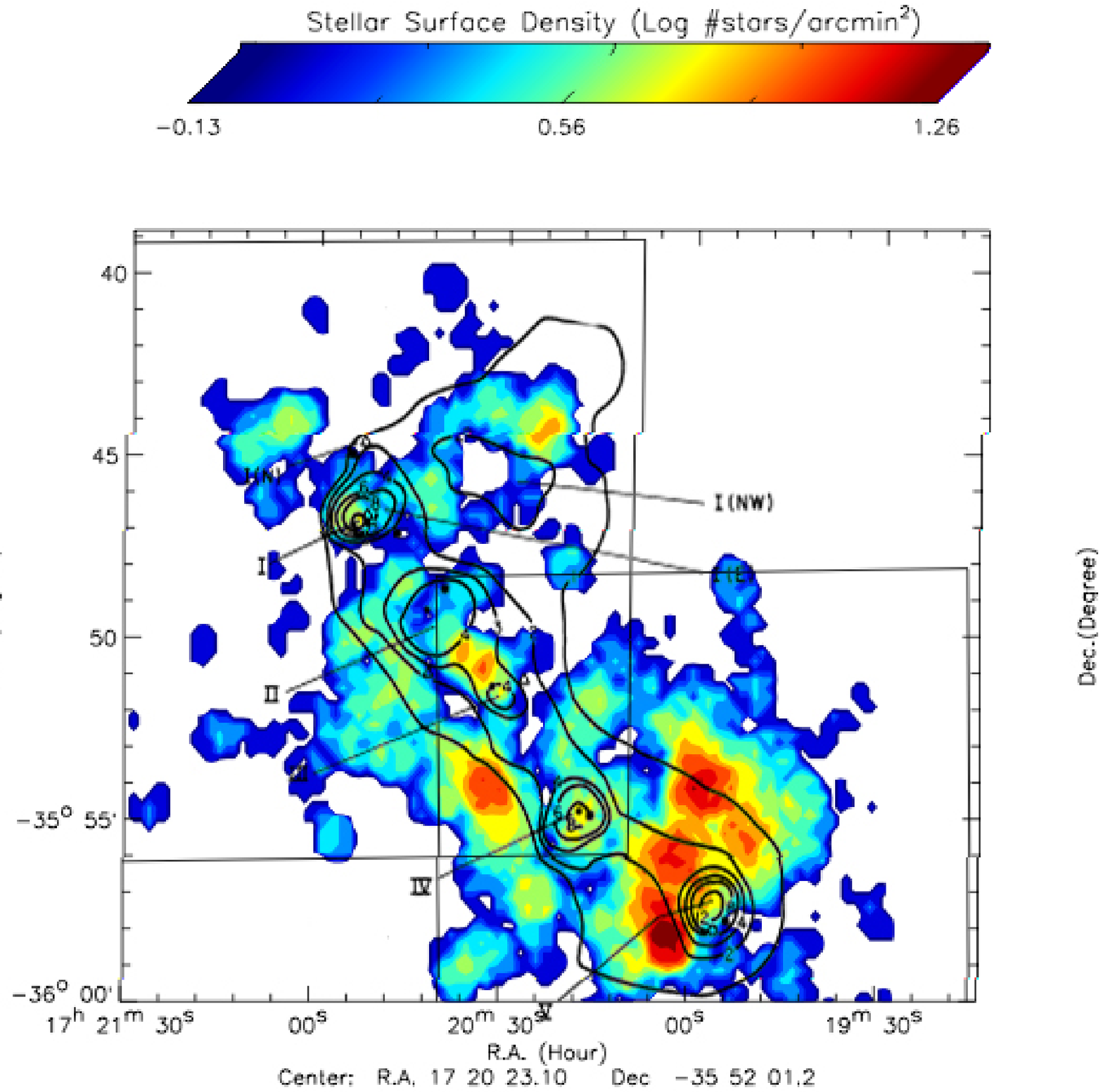}
\includegraphics[width=3in]{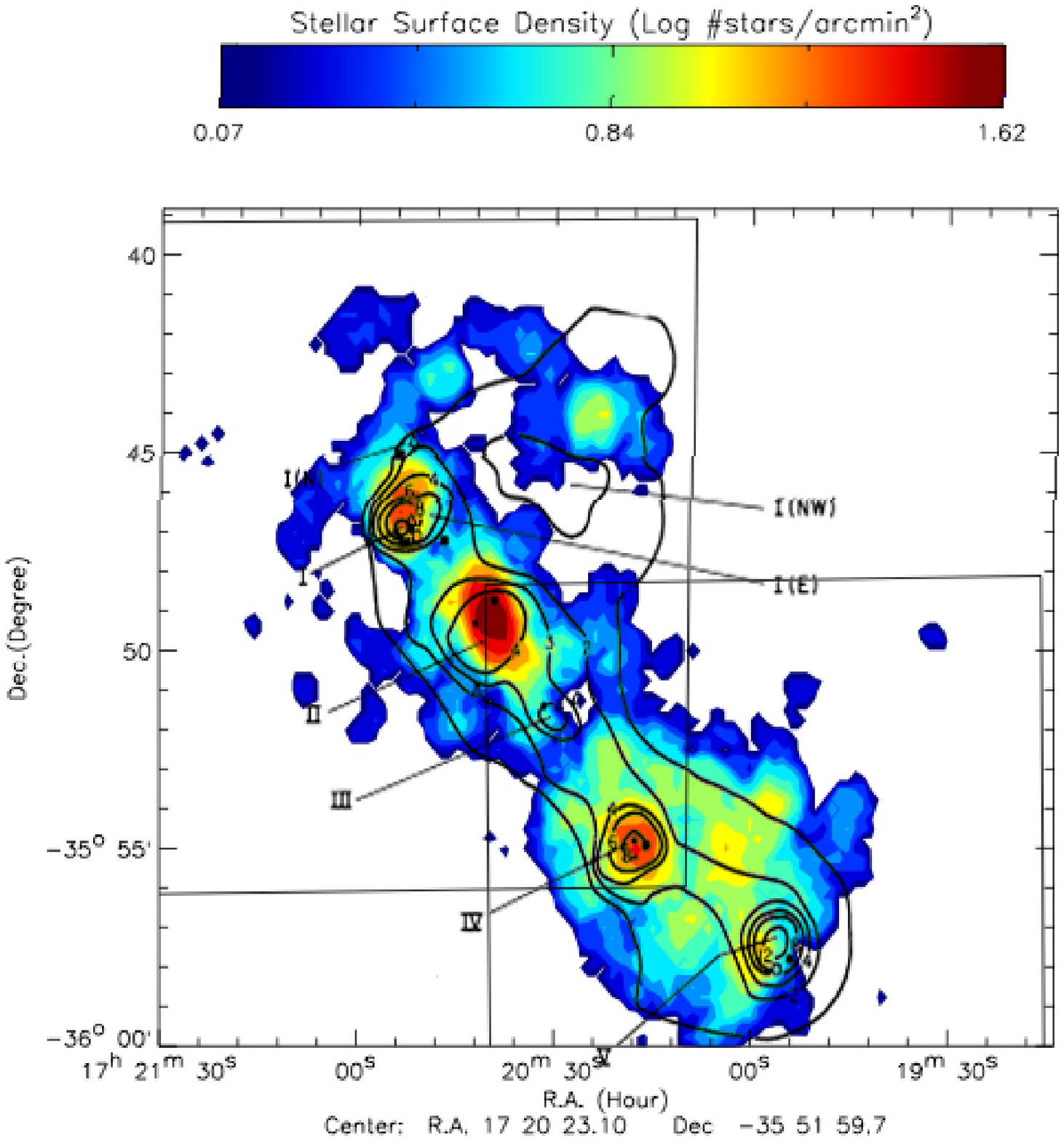}
\caption{The X-ray source map superposed on long-wavelengths maps.  Top: an 18~cm radio continuum map by \citet{Sarma00}.  Bottom: a 71~$\mu$m far-infrared map by \citet{Loughran86}.  The left and right panels show the soft and hard X-ray stellar surface density maps, respectively. }
\label{fig:overlay}
\end{center}
\end{figure}

\begin{figure}[htbp]
\begin{center}
\includegraphics[width=6in]{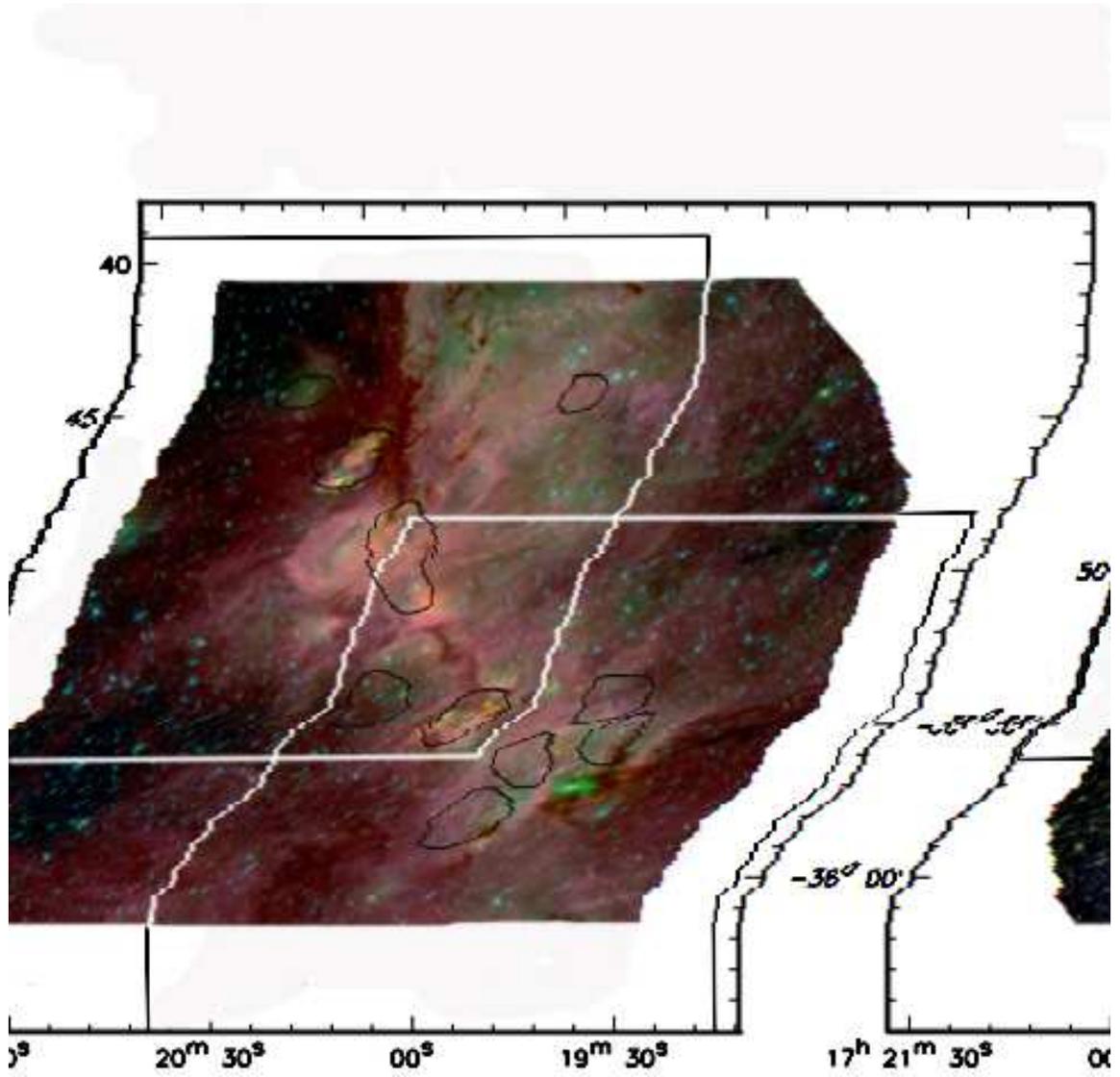}
\caption{Spitzer GLIMPSE map with the X-ray fields and cluster superposed.  Here the 3.6$\mu$m, 5.6$\mu$m, and 8$\mu$m intensities are represented as blue, green and red, respectively. \ref{fig:GLIMPSE} }
\label{fig:GLIMPSE}
\end{center}
\end{figure}

\newpage

\begin{deluxetable}{cccccccccccrcll}
\centering \rotate \tabletypesize{\tiny} \tablewidth{0pt}
\tablecolumns{15}

\tablecaption{{\em Chandra} Catalog of X-ray Sources in NGC 6334 \label{tbl:catalog_stub}}

\tablehead{
\multicolumn{2}{c}{Source} &
\multicolumn{4}{c}{Position} &
\multicolumn{7}{c}{X-ray Properties} &
\multicolumn{2}{c}{Star}  \\
                                
\multicolumn{2}{c}{\hrulefill} &  
\multicolumn{4}{c}{\hrulefill} &
\multicolumn{7}{c}{\hrulefill} &
\multicolumn{2}{c}{\hrulefill}   \\ 

\colhead{Seq.} & \colhead{CXOU J} &
\multicolumn{2}{c}{$\alpha$  ~(J2000.0)~  $\delta$} & \colhead{$\sigma$} & \colhead{$\theta$} &
\colhead{$C_{t}$} & \colhead{$\sigma_{t}$} & \colhead{$B_{t}$} & \colhead{$C_{h}$} & \colhead{PSF} &   
\colhead{$\log P_B$} &  \colhead{$E_{med}$}  & \colhead{$K_s$}  & \colhead{Pos. ID}  \\

\colhead{} & \colhead{} &
\colhead{(deg)} & \colhead{(deg)} & \colhead{('')} & \colhead{(')} &
\colhead{(cts)} & \colhead{(cts)} & \colhead{(cts)} & \colhead{(cts)} & \colhead{} &
\colhead{} &  \colhead{(keV)} & \colhead{(mag)}   &  \\

\colhead{(1)} & \colhead{(2)} &
\colhead{(3)} & \colhead{(4)} & \colhead{(5)} & \colhead{(6)} &
\colhead{(7)} & \colhead{(8)} & \colhead{(9)} & \colhead{(10)} & \colhead{(11)} &
\colhead{(12)}  & \colhead{(13)}  & \colhead{(14)} & \colhead{(15)}
 }

\startdata
1424 & 172053.21$-$354305.7 & 260.221741 & -35.718252 &  0.2 &  4.2 &     4.9 &   2.8 &   0.1 &     5.0 & 0.39 & $<$-5.0 &   3.7  & \nodata & SM 6 \\
1425 & 172053.30$-$354943.2 & 260.222091 & -35.828676 &  0.3 &  2.4 &     2.8 &   2.3 &   0.2 &     1.9 & 0.89 &	-3.0    &   2.0  & \nodata & \nodata \\
1426 & 172053.34$-$354136.7 & 260.222287 & -35.693554 &  0.8 &  5.7 &     4.7 &   3.0 &   1.3 &     3.0 & 0.90 &	-3.1    &   3.5  & \nodata & \nodata \\
1427 & 172053.37$-$354329.4 & 260.222395 & -35.724858 &  0.4 &  3.8 &     5.6 &   3.0 &   0.4 &     5.8 & 0.90 & $<$-5.0 &   4.2  & \nodata & \nodata \\
1428 & 172053.41$-$354701.7 & 260.222549 & -35.783810 &  0.1 &  0.3 &    13.8 &   4.3 &   0.2 &  11.8 & 0.71 & $<$-5.0 &   2.8  & 10.39 & IRS-I = IRN I-1 = I(N)-SMA 3  \\
1429 & 172053.44$-$354704.2 & 260.222677 & -35.784503 &  0.2 &  0.3 &     4.6 &   2.8 &   0.4 &     3.6 & 0.83 &    -4.1    &   3.0  & \nodata & \nodata \\
1430 & 172053.48$-$354702.3 & 260.222866 & -35.783987 &  0.1 &  0.3 &     3.8 &   2.5 &   0.2 &     2.8 & 0.62 &   -4.2     &   3.1  & 10.6: & \nodata \\
1431 & 172053.55$-$354704.6 & 260.223137 & -35.784620 &  0.1 &  0.2 &     9.0 &   3.7 &   1.0 &     9.0 & 0.51 & $<$-5.0 &   4.1  & \nodata & \nodata \\
1432 & 172053.56$-$355029.2 & 260.223199 & -35.841458 &  2.3 &  5.4 &     1.0 &   4.4 &  13.0 &    0.0 & 0.90 & $<$-5.0 &   1.2  & 12.79 & \nodata \\
1433 & 172053.57$-$354600.7 & 260.223221 & -35.766872 &  0.2 &  1.3 &     2.8 &   2.3 &   0.2 &     2.8 & 0.90 &   -3.3     &   5.4  & \nodata & \nodata \\
1434 & 172053.58$-$354636.0 & 260.223279 & -35.776668 &  0.2 &  0.7 &     4.8 &   2.8 &   0.2 &     2.9 & 0.90 & $<$-5.0 &   2.1  & 12.79 & FIR-I 18 = TPR 117  \\
1435 & 172053.60$-$354704.3 & 260.223339 & -35.784538 &  0.1 &  0.3 &     5.6 &   3.0 &   0.4 &     5.6 & 0.41 &   -4.9     &   4.3  & \nodata & \nodata \\
1436 & 172053.64$-$354548.5 & 260.223502 & -35.763493 &  0.1 &  1.5 &     8.0 &   3.4 &   0.0 &     8.0 & 0.41 & $<$-5.0 &   4.1  & \nodata & \nodata \\
1437 & 172053.65$-$354650.0 & 260.223553 & -35.780566 &  0.2 &  0.5 &     3.8 &   2.5 &   0.2 &     3.8 & 0.90 &   -4.7     &   4.2  & \nodata & TPR 83  \\
1438 & 172053.76$-$354642.4 & 260.224023 & -35.778449 &  0.2 &  0.6 &     3.8 &   2.5 &   0.2 &     3.9 & 0.89 & $<$-5.0 &   3.1  & \nodata & TPR 105 \\
1439 & 172053.79$-$354537.0 & 260.224127 & -35.760301 &  0.2 &  1.7 &     3.8 &   2.5 &   0.2 &     0.8 & 0.90 &   -4.7     &   1.5  & \nodata & TPR 197 \\
1440 & 172053.84$-$354620.0 & 260.224356 & -35.772224 &  0.1 &  1.0 &     3.0 &   2.3 &   0.0 &     2.0 & 0.39 & $<$-5.0 &   2.8  & \nodata & TPR 148 \\
1441 & 172053.97$-$354639.4 & 260.224885 & -35.777635 &  0.2 &  0.7 &     4.8 &   2.8 &   0.2 &     4.9 & 0.89 & $<$-5.0 &   3.3  & 12.6: & FIR-I 16 =  TPR 111 \\
1442 & 172054.04$-$354548.5 & 260.225205 & -35.763475 &  0.2 &  1.5 &     5.8 &   3.0 &   0.2 &     5.8 & 0.90 & $<$-5.0 &   4.5  & \nodata & TPR 185 \\
1443 & 172054.14$-$354605.7 & 260.225603 & -35.768258 &  0.2 &  1.2 &     6.8 &   3.2 &   0.2 &     6.9 & 0.90 & $<$-5.0 &   4.2  & \nodata & TPR 160  \\
\enddata

\tablecomments{Table~\ref{tbl:catalog_stub} is available in its entirety in the electronic edition of the Journal.  Only a portion of the table around NGC~6334 I(N) is shown here. 
 \\Col.\ (1): X-ray catalog sequence number, sorted by RA.
 \\Col.\ (2): IAU designation.
 \\Cols.\ (3) and (4): Right ascension and declination for epoch (J2000.0).
 \\Col.\ (5): Estimated standard deviation of the random component of the position error, $\sqrt{\sigma_x^2 + \sigma_y^2}$.  The single-axis position errors, $\sigma_x$ and $\sigma_y$, are estimated from the single-axis standard deviations of the PSF inside the extraction region and the number of counts extracted.
 \\Col.\ (6): Off-axis angle.
 \\Cols.\ (7) and (8): Net counts extracted in the total energy band (0.5--8~keV); average of the upper and lower $1\sigma$ errors on col.\ (7).
 \\Col.\ (9): Background counts expected in the source extraction region (total band).
 \\Col.\ (10): Net counts extracted in the hard energy band (2--8~keV).
 \\Col.\ (11): Fraction of the PSF (at 1.5 keV) enclosed within the extraction region. A reduced PSF fraction (significantly below 90\%) may indicate that the source is in a crowded region. 
\\Col.\ (12): Log of the probability that the source arises from fluctuations in the local background based on Poisson statistics. 
\\Col.\ (13): Background-corrected median photon energy (total band).
\\Col.\ (14): $K_s$ magnitude of 2MASS Point Source Catalog stars within 1\arcsec\/ of the X-ray position. ``:'' indicates photometric confusion by a nearby star or nebular emission, or photometric accuracy worse than $\pm 0.2$~mag. ``$>$'' indicates source was detected in the $J$ or $H$ bands but not the $K_s$ band. 
\\Col.\ (15): Possible counterparts within 3\arcsec\/ of the X-ray source in the SIMBAD database of published objects. 
FIR-V sources are from \citet{Straw89}, TPR stars are from \citet{Tapia96}, and other designations are discussed in the text.}
 
\end{deluxetable}

\newpage
  
\begin{deluxetable}{ccccccrccc}
\centering
\tabletypesize{\footnotesize}
\tablewidth{0pt}
\tablecolumns{10}

\tablecaption{X-ray Star Clusters in NGC~6334 \label{tbl:xray}}
\tablehead{
\multicolumn{7}{c}{X-ray Clusters} &
\multicolumn{1}{c}{IRAS} &
\multicolumn{1}{c}{FIR} &
\multicolumn{1}{c}{Radio} \\

\multicolumn{7}{c}{\hrulefill} &  
\multicolumn{1}{c}{\hrulefill} &
\multicolumn{1}{c}{\hrulefill} &
\multicolumn{1}{c}{\hrulefill} \\

\colhead{ID} & \colhead{$\alpha$} & \colhead{$\delta$} & \colhead{Soft} & \colhead{Hard} &\colhead{Extent} & \colhead{$N_x$} & \colhead{} &\colhead{} &\colhead{}  \\

\colhead{(1)} & \colhead{(2)} &
\colhead{(3)} & \colhead{(4)} & \colhead{(5)} & \colhead{(6)} &\colhead{(7)} &
\colhead{(8)} & \colhead{(9)}& \colhead{(10)}}

\startdata
XA      &  17:19:54 &   -35:55:30 & Y & N  & $2^{\prime}\times 2^{\prime}$ &   ~9 &    \nodata     & \nodata & \nodata   \\
XB      &  17:19:58 &   -35:54:00 & Y & W & $3^{\prime}\times 3^{\prime}$ &  ~31 &   \nodata     & \nodata & \nodata   \\
XC      &  17:20:02 &   -35:56:00 & Y & W & $2^{\prime}\times 2^{\prime}$ &  ~36 &nr 17165-3554 & \nodata & \nodata   \\
XD      &  17:20:04 &   -35:58:15 & Y & W & $1^{\prime}\times 1^{\prime}$ &  ~43 &   \nodata     & V           & \nodata   \\
XE      &  17:20:18 &   -35:55:00 & W & Y & $2^{\prime}\times 2^{\prime}$ &   ~48 &   \nodata    & IV          &   A           \\
XF      &  17:20:24 &   -35:44:15 & Y  & W & $1^{\prime}\times 1^{\prime}$ &     ~7 &   \nodata    & \nodata  & \nodata   \\
XG      &  17:20:33 &  -35:54:00 & Y & N & $2^{\prime}\times 2^{\prime}$ &    ~25 &   \nodata    & \nodata  & \nodata   \\
XH      &  17:20:38 &  -35:49:00 & Y & W & $3^{\prime}\times 4^{\prime}$ &    ~94 &17172-3548 & II           &   D           \\
XI       &   17:20:53 &  -35:46:00 & N & Y & $1^{\prime}\times 2^{\prime}$ &    ~15 &17175-3544 & I, I(N)    &   E, F     \\
XJ      &   17:21:05 &  -35:44:00 & Y & N & $1^{\prime}\times 1^{\prime}$ &     ~5 &   \nodata     & \nodata  &  \nodata  \\
\enddata

\tablecomments{
\\Col.\ (1): X-ray Cluster designation
\\Col.\ (2) and (3): Right ascension and declination for epoch (J2000.0)
\\Col.\ (4)and (5): Indicator of cluster richness in soft (col. 4) and hard (col. 5) band: Y = Yes, N = No, W = Weak
\\Col.\ (6): The spatial extent of the x-ray cluster, rounded to the nearest arcminute.
\\Col.\ (7): X-ray source count in the X-ray cluster.
\\Cols.\ (8): IRAS point source catalog counterpart.  nr = near.
\\Cols.\ (9): Far-infrared counterpart \citep{Loughran86}
\\Cols.\ (10): Radio HII region counterpart \citep{Rodriguez82}.}

\end{deluxetable}



\newpage

\begin{thebibliography}

\bibitem[Albacete Colombo et al.(2007)]{Albacete07} Albacete Colombo, J.~F., Caramazza, M., Flaccomio, E., Micela, G., \& Sciortino, S.\ 2007, \aap, 474, 495 
\bibitem[Baraffe et al.(1998)]{Baraffe98} Baraffe, I., Chabrier, G., Allard, F., \& Hauschildt, P.~H.\ 1998, \aap, 337, 403
\bibitem[Bassani et al.(2005)]{Bassani05} Bassani, L., et al.\  2005, \apjl, 634, L21
\bibitem[Benjamin et al.(2003)]{Benjamin03} Benjamin, R.~A., et al.\ 2003, \pasp, 115, 953  
\bibitem[Bessell \& Brett(1988)]{Bessell88} Bessell, M.~S., \& Brett, J.~M.\ 1988, \pasp, 100, 1134 
\bibitem[Beuther et al.(2007)]{Beuther07} Beuther, H., Walsh, A.~J., Thorwirth, S., Zhang, Q., Hunter, T.~R., Megeath, S.~T., \& Menten, K.~M.\ 2007, \aap, 466, 989
\bibitem[Beuther et al.(2008)]{Beuther08} Beuther, H., Walsh, A.~J., Thorwirth, S., Zhang, Q., Hunter, T.~R., Megeath, S.~T., \& Menten, K.~M.\ 2008, \aap, 481, 169 
\bibitem[Broos et al.(2007)]{Broos07} Broos, P.~S., Feigelson, E.~D., Townsley, L.~K., Getman, K.~V., Wang, J., Garmire, G.~P., Jiang, Z. \& Tsuboi, Y.\ 2007, \apjs, 169, 353 
\bibitem[Burton et al.(2000)]{Burton00} Burton, M.~G., et al.\ 2000, \apj, 542, 359 
\bibitem[Bykov et al.(2006)]{Bykov06} Bykov, A.~M., et al.\  2006, \apjl, 649, L21 
\bibitem[Carey et al.(1998)]{Carey98} Carey, S.~J., Clark, F.~O., Egan, M.~P., Price, S.~D., Shipman, R.~F., 
\& Kuchar, T.~A.\ 1998, \apj, 508, 721 
\bibitem[Carral et al.(2002)]{Carral02} Carral, P., Kurtz, S.~E., Rodr{\'{\i}}guez, L.~F., Menten, K., Cant{\'o}, J., 
\& Arceo, R.\ 2002, \aj, 123, 2574  
\bibitem[Chabrier(2003)]{Chabrier03} Chabrier, G.\ 2003, \pasp, 115, 763
\bibitem[Chlebowski et al.(1989)]{Chlebowski89} Chlebowski, T., Harnden, F.~R., Jr., \& Sciortino, S.\ 1989, \apj, 341, 427 
\bibitem[Cohen et al.(2003)]{Cohen03} Cohen, D.~H., de 
Messi{\`e}res, G.~E., MacFarlane, J.~J., Miller, N.~A., Cassinelli, J.~P., 
Owocki, S.~P., \& Liedahl, D.~A.\ 2003, \apj, 586, 495 
\bibitem[De Buizer et al.(2002)]{deBuizer02} De Buizer, J.~M., Radomski, J.~T., Pi{\~n}a, R.~K., \& Telesco, C.~M.\ 2002, \apj, 580, 305 
\bibitem[Ezoe et al.(2006)]{Ezoe06} Ezoe, Y., Kokubun, M., Makishima, K., Sekimoto, Y., \& Matsuzaki, K.\ 2006, \apj, 638, 860
\bibitem[Feigelson et al.(2002)]{Feigelson02} Feigelson, E.~D., Broos, P., Gaffney, J.~A., III, Garmire, G., Hillenbrand, L.~A., Pravdo, S.~H., Townsley, L., \& Tsuboi, Y.\ 2002, \apj, 574, 258
\bibitem[Feigelson et al.(2005)]{Feigelson05} Feigelson, E.~D., et al.\ 2005, \apjs, 160, 379 
\bibitem[Feigelson et al.(2007)]{Feigelson07} Feigelson, E., Townsley, L., G{\"u}del, M., \& Stassun, K.\ 2007, Protostars and Planets V, 313 
\bibitem[Feigelson \& Townsley(2008)]{Feigelson08} Feigelson, E.~D., \& Townsley, L.~K.\ 2008, \apj, 673, 354 
\bibitem[Freeman et al.(2002)]{Freeman02} Freeman, P.~E., Kashyap, V., Rosner, R., \& Lamb, D.~Q.\ 2002, \apjs, 138, 185 
\bibitem[Gagn{\'e} et al.(2005)]{Gagne05} Gagn{\'e}, M., Oksala, M.~E., Cohen, D.~H., Tonnesen, S.~K., ud-Doula, A., Owocki, S.~P., Townsend, R.~H.~D., \& MacFarlane, J.~J.\ 2005, \apj, 628, 986 
\bibitem[Garmire et al.(2003)]{Garmire03} Garmire, G.~P., Bautz, 
M.~W., Ford, P.~G., Nousek, J.~A., G.~R., Jr.\ 2003, \procspie, 4851, 28 
\bibitem[Getman et al.(2005)]{Getman05} Getman, K.~V., et al.\ 2005, \apjs, 160, 319 
\bibitem[Getman et al.(2006)]{Getman06} Getman, K.~V., Feigelson, E.~D., Townsley, L., Broos, P., Garmire, G., \& Tsujimoto, M.\ 2006, \apjs, 163, 306 
\bibitem[Getman et al.(2007)]{Getman07} Getman, K.~V., Feigelson, E.~D., Garmire, G., Broos, P., \& Wang, J.\ 2007, \apj, 654, 316 
\bibitem[Gorenstein(1975)]{Gorenstein75} Gorenstein, P.\ 1975, \apj, 198, 95 
\bibitem[Grosso et al.(2005)]{Grosso05} Grosso, N., et al.\ 2005, \apjs, 160, 530 
\bibitem[G{\"u}del et al.(2007)]{Gudel07} G{\"u}del, M., et al.\ 2007, \aap, 468, 353 
\bibitem[Hamaguchi et al.(2005)]{Hamaguchi05} Hamaguchi, K., Corcoran, M.~F., Petre, R., White, N.~E., Stelzer, B., Nedachi, K., Kobayashi, N., \& Tokunaga, A.~T.\ 2005, \apj, 623, 291 
\bibitem[Hashimoto et al.(2008)]{Hashimoto08} Hashimoto, J., et al.\ 2008, \apjl, 677, L39 
\bibitem[Hunter et al.(2006)]{Hunter06} Hunter, T.~R., Brogan, C.~L., Megeath, S.~T., Menten, K.~M., Beuther, H., \& Thorwirth, S.\ 2006, \apj, 649, 888 
\bibitem[Jackson \& Kraemer(1999)]{Jackson99} Jackson, J.~M., \& Kraemer, K.~E.\ 1999, \apj, 512, 260 
\bibitem[Kraemer et al.(1999)]{Kraemer99} Kraemer, K.~E., Deutsch, L.~K., Jackson, J.~M., Hora, J.~L., Fazio, G.~G., Hoffmann, W.~F., \& Dayal, A.\ 1999, \apj, 516, 817 
\bibitem[Lada \& Adams(1992)]{Lada92} Lada, C.~J., \& Adams, F.~C.\ 1992, \apj, 393, 278
\bibitem[Lindsay(1955)]{Lindsay55} Lindsay, E.~M.\ 1955, Irish Astronomical Journal, 3, 182 
\bibitem[Loughran et al.(1986)]{Loughran86} Loughran, L., McBreen, B., Fazio, G.~G., Rengarajan, T.~N., Maxson, C.~W., Serio, S., Sciortino, S., \& Ray, T.~P.\ 1986, \apj, 303, 629 
\bibitem[Lucy(1974)]{Lucy74} Lucy, L.~B.\ 1974, \aj, 79, 745 
\bibitem[Matthews et al.(2008)]{Matthews08} Matthews, H.~E., McCutcheon, W.~H., Kirk, H., White, G.~J., \& Cohen, M.\ 2008, \aj, 136, 2083 
\bibitem[McBreen et al.(1983)]{McBreen83} McBreen, B., Jaffe, D. T., \& Fazio, G. G. 1983,  \aj, 88, 835
\bibitem[Meyer et al.(1997)]{Meyer97} Meyer, M.~R., Calvet, N., \& Hillenbrand, L.~A.\ 1997, \aj, 114, 288 
\bibitem[Moran \& Rodriguez(1980)]{Moran80} Moran, J.~M., \& Rodriguez, L.~F.\ 1980, \apjl, 236, L159 
\bibitem[Mori et al.(2001)]{Mori01} Mori, K., Tsunemi, H., Miyata, E., Baluta, C.~J., Burrows, D.~N., Garmire, G.~P., \& Chartas, G.\ 2001, New Century of X-ray Astronomy, 251, 576 
\bibitem[Mu{\~n}oz et al.(2007)]{Munoz07} Mu{\~n}oz, D.~J., Mardones, D., Garay, G., Rebolledo, D., Brooks, K., \& Bontemps, S.\ 2007, \apj, 668, 906 
\bibitem[Neckel(1978)]{Neckel78} Neckel, T. \ 1978, \aap, 69, 51
\bibitem[Persi et al.(2000)]{Persi00} Persi, P., Tapia, M., \& Roth, M.\ 2000, \aap, 357, 1020 
\bibitem[Persi \& Tapia (2008)]{Persi08}  Persi, P., \& Tapia, M. \ 2008, in Handbook of Star Forming Regions,  vol II,  San Francisco: Astro.\ Soc.\ Pacific, 456
\bibitem[Persi et al.(2009)]{Persi09} Persi, P., Tapia, M., Roth, M., \& G{\'o}mez, M.\ 2009, \aap, 493, 571
\bibitem[Rathborne et al.(2006)]{Rathborne06} Rathborne, J.~M., Jackson, J.~M., \& Simon, R.\ 2006, \apj, 641, 389
\bibitem[Rho et al.(2004)]{Rho04} Rho, J., Ram{\'{\i}}rez, S.~V., Corcoran, M.~F., Hamaguchi, K., \& Lefloch, B.\ 2004, \apj, 607, 904 
\bibitem[Rodriguez et al.(1982)]{Rodriguez82} Rodriguez, L.~F., Canto, J., \& Moran, J.~M.\ 1982, \apj, 255, 103 
\bibitem[Rodriguez et al.(1988)]{Rodriguez88} Rodriguez, L.~F., Canto, J., \& Moran, J.~M.\ 1988, \apj, 333, 801 
\bibitem[Rom{\'a}n-Z{\'u}{\~n}iga et al.(2008)]{Roman08} Rom{\'a}n-Z{\'u}{\~n}iga, C.~G., Elston, R., Ferreira, B., 
\& Lada, E.~A.\ 2008, \apj, 672, 861 
\bibitem[Sana et al.(2006)]{Sana06} Sana, H., Rauw, G., 
Naz{\'e}, Y., Gosset, E., \& Vreux, J.-M.\ 2006, \mnras, 372, 661 
\bibitem[Sandell(1999)]{Sandell99} Sandell, G.\ 1999, \aap, 343, 281 
\bibitem[Sandell(2000)]{Sandell00} Sandell, G.\ 2000, \aap, 358, 242
\bibitem[Sarma et al.(2000)]{Sarma00} Sarma, A.~P., Troland, T.~H., Roberts, D.~A., \& Crutcher, R.~M.\ 2000, \apj, 533, 271 
\bibitem[Schulz et al.(2003)]{Schulz03} Schulz, N.~S., 
Canizares, C., Huenemoerder, D., \& Tibbets, K.\ 2003, \apj, 595, 365 
\bibitem[Seifahrt et al.(2008)]{Seifahrt08} Seifahrt, A., Thorwirth, S., Beuther, H., Leurini, S., Brogan, C.~L., Hunter, T.~R., Menten, K.~M., \& Stecklum, B.\ 2008, J.\ Physics Conf.\ Ser., 131, 012030 
\bibitem[Sekimoto et al.(2000)]{Sekimoto00} Sekimoto, Y.,  Matsuzaki, K., Kamae, T., Tatematsu, K., Yamamoto, S., \& Umemoto, T.\ 2000, \pasj, 52, L31 
\bibitem[Sharpless(1959)]{Sharpless59} Sharpless, S.\ 1959, \apjs, 4, 257 
\bibitem[Siess et al.(2000)]{Siess00} Siess, L., Dufour, E., \& Forestini, M.\ 2000, \aap, 358, 593 
\bibitem[Skinner et al.(2005)]{Skinner05} Skinner, S.~L., Zhekov, 
S.~A., Palla, F., \& Barbosa, C.~L.~D.~R.\ 2005, \mnras, 361, 191 
\bibitem[Stelzer et al.(2006)]{Stelzer06} Stelzer, B., Hu{\'e}lamo, N., Micela, G., \& Hubrig, S.\ 2006, \aap, 452, 1001 
\bibitem[Straw \& Hyland(1989)]{Straw89} Straw, S.~M., \& Hyland, A.~R.\ 1989, \apj, 340, 318
\bibitem[Straw et al.(1989)]{Straw89b} Straw, S.~M., Hyland, A.~R., \& McGregor, P.~J.\ 1989, \apjs, 69, 99  
\bibitem[Tapia et al.(1996)]{Tapia96} Tapia, M., Persi, P., \& Roth, M.\ 1996, \aap, 316, 102 
\bibitem[Telleschi et al.(2007)]{Telleschi07} Telleschi, A., G{\"u}del, M., Briggs, K.~R., Audard, M., \& Palla, F.\ 2007, \aap, 468, 425 
\bibitem[Townsley et al.(2003)]{Townsley03} Townsley, L.~K., Feigelson, E.~D., Montmerle, T., Broos, P.~S., Chu, Y.-H., \& Garmire, G.~P. \ 2003, \apj, 593, 874 
\bibitem[Townsley et al.(2006)]{Townsley06a} Townsley, L.~K., Broos, P.~S., Feigelson, E.~D., Garmire, G.~P., \& Getman, K.~V.\ 2006, \aj, 131, 2164 
\bibitem[Trotter et al.(1998)]{Trotter98} Trotter, A.~S., Moran, J.~M., \& Rodriguez, L.~F.\ 1998, \apj, 493, 666 
\bibitem[Vuong et al.(2003)]{Vuong03} Vuong, M.~H., Montmerle, T., Grosso, N., Feigelson, E.~D., Verstraete, L., \& Ozawa, H.\ 2003, \aap, 408, 581 
\bibitem[Walborn(1982)]{Walborn82} Walborn, N.~R.\ 1982, \aj, 87, 1300 
\bibitem[Wang et al.(2007)]{Wang07} Wang, J., Townsley, L.~K., Feigelson, E.~D., Getman, K.~V., Broos, P.~S., Garmire, G.~P., \& Tsujimoto, M.\ 2007, \apjs, 168, 100 
\bibitem[Wang et al.(2008)]{Wang08} Wang, J., Townsley, L.~K., Feigelson, E.~D., Broos, P.~S., Getman, K.~V., Rom{\'a}n-Z{\'u}{\~n}iga, C.~G., \& Lada, E.\ 2008, \apj, 675, 464 
\bibitem[Wang et al.(2009)]{Wang09} Wang, J., Townsley, L.~K., Feigelson, E.~D., Broos, P.~S., Getman, K.~V., Rom{\'a}n-Z{\'u}{\~n}iga, C.~G., \& Lada, E.\ 2009, \apj, in press 
\bibitem[Weisskopf et al.(2002)]{Weisskopf02} Weisskopf, M.~C., Brinkman, B., Canizares, C., Garmire, G., Murray, S., \& Van Speybroeck, L.~P.\ 2002, \pasp, 114, 1 
\bibitem[Weisskopf et al.(2007)]{Weisskopf07} Weisskopf, M.~C., Wu, K., Trimble, V., O'Dell, S.~L., Elsner, R.~F., Zavlin, V.~E., \& Kouveliotou, C.\ 2007, \apj, 657, 1026 
\bibitem[Wolk et al.(2006)]{Wolk06} Wolk, S.~J., Spitzbart, B.~D., Bourke, T.~L., \& Alves, J.\ 2006, \aj, 132, 1100 
 \end{thebibliography}
\end{document}